\documentclass[transmag]{IEEEtran}
\usepackage{latexsym}
\usepackage{graphicx}
\usepackage{amsfonts,amssymb,amsmath}
\usepackage{hyperref}
\usepackage{xcolor}
\usepackage{cite}
\usepackage[utf8]{inputenc}
\usepackage[english]{babel}
\def\BibTeX{{\rm B\kern-.05em{\sc i\kern-.025em b}\kern-.08em T\kern-.1667em\lower.7ex\hbox{E}\kern-.125emX}}
\begin{document}

\title{\textcolor{black}{On} Integrated Access and Backhaul Networks: Current Status and Potentials}

\author{Charitha Madapatha, Behrooz Makki,
\IEEEmembership{Senior Member, IEEE,} Chao Fang, Oumer Teyeb, Erik Dahlman, 
\\
Mohamed-Slim Alouini, 
\IEEEmembership{Fellow, IEEE}
and Tommy Svensson,
\IEEEmembership{Senior Member, IEEE.}
\thanks{Manuscript received xxxx; accepted November xxxx. Date of publication xxxx; date of current version xxxx.This work was supported by the ChaseOn project of Dept. of Electrical Engineering, Chalmers University of Technology. The work of C. Madapatha in this publication is part of his research work at Chalmers University of Technology, funded by a Swedish Institute scholarship.}
\thanks{C. Madapatha, C. Fang and T. Svensson are with the Department of Electrical Engineering, Chalmers University of Technology, 412 96 Gothenburg, Sweden (e-mail: charitha@student.chalmers.se; fchao@chalmers.se; tommy.svensson@chalmers.se).}
\thanks{B. Makki is with Ericsson research, Lindholmen, Sweden, 417 56 Göteborg, Sweden (e-mail: behrooz.makki@ericsson.com).}
\thanks{O. Teyeb and E. Dahlman are with Ericsson research, Kista, Sweden, P.O. Box 16440 (e-mail: {oumer.teyeb, erik.dahlman}@ericsson.com).}
\thanks{M.-S. Alouini is with the Computer, Electrical and Mathematical Science and Engineering, King Abdullah University of Science and Technology (KAUST), Thuwal 23955-6900, Saudi Arabia (e-mail: slim.alouini@kaust.edu.sa).}}

\IEEEtitleabstractindextext{\begin{abstract}
—In this paper, we introduce and study the potentials and challenges of integrated access and backhaul (IAB) \textcolor{black}{as one of the promising techniques for evolving 5G networks}. We study IAB networks from different perspectives. We summarize the recent Rel-16 as well as the upcoming Rel-17 3GPP discussions on IAB, and highlight the main IAB-specific agreements on different \textcolor{black}{protocol} layers. Also, concentrating on millimeter wave-based communications, we evaluate the performance of IAB networks in both dense and \textcolor{black}{suburban} areas. Using a finite stochastic geometry model, with random distributions of IAB nodes as well as user equipments (UEs) in a finite region, we study the service coverage rate defined as the probability of the event that the UEs' minimum rate requirements are satisfied. We present comparisons between IAB \textcolor{black}{and hybrid IAB/fiber-backhauled networks} where \textcolor{black}{a part or all of} the small \textcolor{black}{base stations} are fiber-connected. Finally, we study the robustness of IAB networks to \textcolor{black}{weather and various deployment conditions} and verify \textcolor{black}{their effects, such as blockage, tree foliage, rain as well as antenna height/gain} on the coverage rate of IAB setups, as the key differences between the fiber-connected and IAB networks. As we show, IAB is an attractive \textcolor{black}{approach to enable the network densification required by 5G and beyond.} 

\end{abstract}


}

\maketitle
\section{INTRODUCTION}
Different reports, e.g., \cite{refi1}, predict a steep increase of Internet devices connected through wireless access as well as a massive increase in mobile traffic. To cope with such requirements, \textcolor{black}{along with utilizing more spectrum,} the fifth generation (5G) wireless networks \textcolor{black}{and beyond} propose different ways for spectral efficiency \textcolor{black}{and capacity improvements. Network densification \cite{MS0}, \cite{MS1} is one of the key enablers among the alternative approaches, e.g., various distributed antenna systems techniques, including cell-free massive multiple-input-multiple-output (MIMO)} and can be achieved via the deployment of many \textcolor{black}{access points} of different types, so that there are more resource blocks per unit area. 


\textcolor{black}{The base stations (BSs) \textcolor{black}{need to be} connected to the operators’ core network via a transport network. A transport network may consist of wired or wireless connections. Typically, wireless connections are used for backhaul transport in the radio access network (RAN), closer to the BSs, while wired high-capacity fiber connections are used for transport closer to the core \textcolor{black}{network} and in the core network, where the network needs to handle aggregated traffic from many BSs.  }

\textcolor{black}{The deployed backhaul technology \textcolor{black}{today} has large regional variations, but on a global scale, wireless microwave technology has historically been a dominating media for a long time. Over the last 10 years there has however been a large increase in fiber deployments attributed to, e.g., geopolitical decisions and major governmental investments. Over the same time, the use of copper as a media has reduced a lot due to increasing demands on capacity and lower maintenance. Going forward there are thus two dominating backhaul media – microwave and fiber. Historical and predicted global backhaul media distribution can be found in \cite{refi1erc}.}

\textcolor{black}{Fiber offers reliable high-capacity transport with demonstrated Tbps rates. However, the deployment of fiber requires a noteworthy initial investment for trenching and installation, \textcolor{black}{which} could take a considerable installation time, and even \textcolor{black}{might not be} possible/allowed in, certain areas where trenching is not an option.}

\textcolor{black}{Wireless backhaul\textcolor{black}{ing} \textcolor{black}{using microwave represents a competitive alternative to fiber since it today provides 10’s of Gbps in commercial deployments and even 100 Gbps has recently been demonstrated \cite{Xref}. Microwave} is a backhaul technology used by most mobile operators worldwide, and this trend is likely to continue. This is because microwave is a scalable and economical backhaul option \textcolor{black}{that can meet} the increasing requirements of 5G systems.} \textcolor{black}{A key advantage over fiber is that wireless backhauling comes} with significantly lower cost and flexible/timely deployment (e.g., no digging, no \textcolor{black}{intrusion or }disruption of infrastructure, and possible to deploy \textcolor{black}{in principle } everywhere) \textcolor{black}{\cite{refi1erc}, \cite{MS2}}. 
\textcolor{black}{Today microwave backhauling} operates in licensed point-to-point (PtP) spectrum, typically in the 4–70/80 GHz range. \textcolor{black}{However,} with the introduction of 5G in millimeter wave (mmWave) spectrum and \textcolor{black}{with} the foreseen need for even wider bandwidths for backhaul, microwave is \textcolor{black}{currently being extended} to even higher frequencies, above 100 GHz.

\textcolor{black}{For the same reasons, and driven by network densification and access to wide bandwidth in mmWave} spectrum, integrated access and backhaul (IAB) networks, where the operator can utilize part of the radio resources for wireless backhauling, has recently received considerable attention \cite{refi3}, \cite{refi4}. The purpose of IAB is to \textcolor{black}{provide} flexible wireless \textcolor{black}{backhauling} using 3GPP \textcolor{black}{new radio (NR)}  technology in \textcolor{black}{ international mobile telecommunications (IMT) bands, \textit{providing not only backhaul but also the existing cellular services in the same node.}}  \textcolor{black}{Thus,} IAB serves as a complement to microwave PtP backhaul\textcolor{black}{ing} in dense urban and suburban deployments, \textcolor{black}{while} it comes at the expense of using IMT bands not only for access but also for backhaul traffic.

\textcolor{black}{Wireless backhauling} has been studied earlier in 3GPP in the scope of LTE Rel-10, also known as LTE relaying \cite{refi5}. However, there have been only a handful of commercial LTE relay deployments, mainly because the existing LTE spectrum is very expensive to be used for backhauling, and also small-cell deployments did not reach the anticipated potential in the 4G timeline.

For 5G \textcolor{black}{NR}, IAB has been standardized in 3GPP Rel-16 \textcolor{black}{and, as we detail later in the paper, standardization will continue in Rel-17.} The main reason why NR IAB is expected to be more commercially successful than LTE relaying \textcolor{black}{is that:
\begin{itemize}
    \item The limited coverage \textcolor{black}{of} mmWave access creates a high demand for denser deployments, which, in turn, increases the need for backhauling.
    \item Also, the larger bandwidth available in mmWave spectrum provides more economically viable opportunity for \textcolor{black}{wireless} backhauling. 
    \item Finally, MIMO, \textcolor{black}{multi-beam systems, and multiple access, which are inherent features of NR, \textcolor{black}{enable} efficient backhauling of multiple \textcolor{black}{radio BSs} using the same equipment.}
\end{itemize}
}

\textcolor{black}{
There have been several studies on the performance of IAB networks. For instance, cost-optimal node placement \cite{refi6}, resource allocation \cite{refi7,refi8,refi9,refi10} and routing \cite{refi7,refi8,refi9,refi10,refi11,refi12, refin1} are studied in the cases with different numbers of hops. Particularly, \cite{refi6} provides \textcolor{black}{an overview of multi-hop IAB} \textcolor{black}{techniques supported} in \textcolor{black}{the} 3GPP rel-16 standard and discusses its design strategies. A joint node placement and resource allocation scheme maximizing the downlink sum rate is developed in \cite{refia}. Also, \cite{refib} uses  simulated annealing algorithms for joint scheduling and power allocation. The maximum extended coverage area of a single fiber site using multi-hop relaying is investigated in \cite{refi13}, and \cite{refi14,refi14x} perform end-to-end simulations to check the feasibility/challenges of mmWave-based IAB networks. \textcolor{black}{Also, \cite{gomez2016optimal} provides useful insights for IAB deployments, especially, related to network densification and multi-hop topology, as it simulates a multi-hop mmWave pico-cell network, and evaluates the user throughput.} The potential of using IAB in a fixed wireless access use-case is evaluated in  \cite{refmikpaper}. The impact of dynamic time division duplex (TDD)-based resource allocation on the throughput of IAB networks, how its performance compares with static TDD and FDD (F: frequency), is discussed in \cite{refi15,refi16}. Moreover, \cite{refi17,refi18,refi19} characterize the coverage probability of IAB-enabled mmWave heterogeneous networks via infinite Poisson point processes (PPPs). Precoder design and power allocation, to maximize the network sum rate, is considered in \cite{refi20}. Finally, \cite{refi20b} investigates the usefulness of IAB in unmanned aerial vehicle (UAV)- based communications, and \cite{refinx} develops a reinforcement learning-based resource allocation \textcolor{black}{scheme} in such networks.
}

\textcolor{black}{
In this paper, we study the performance of IAB networks from different perspectives. We start by summarizing the most recent 3GPP discussions in Rel-16 as well as the upcoming ones in Rel-17, and highlight the main IAB-specific features on different protocol layers. Then, concentrating on mmWave-based communications,  we analyze the performance of IAB networks, and compare their performance with those achieved with hybrid IAB/fiber-connected networks. Here, the results are presented for the cases with an FHPPP (FH: finite homogeneous)-based stochastic geometry model, e.g., \cite{refi17}, \cite{refnewb}, i.e., a \textcolor{black}{PPP} which depends on a constant density, with random distributions of the IAB nodes as well as the user equipments (UEs) in a finite region. Particularly, we study the network service coverage probability, defined as the probability of the event that the UEs' minimum data rate requirements are satisfied.}

One of the key differences \textcolor{black}{between} fiber-connected and IAB networks is that, \textcolor{black}{the backhaul link in IAB networks} \textcolor{black}{may, like any wireless link,} be \textcolor{black}{impacted} by \textcolor{black}{various weather effects and deployment conditions} such as rain, blockage, antenna heights, and tree foliage. For this reason, we evaluate the \textcolor{black}{impacts of these aspects.} \textcolor{black}{The }results are presented for both
\textcolor{black}{suburban} and urban areas, \textcolor{black}{with the main focus} on dense deployments, since that is the most interesting scenario for \textcolor{black}{IAB.}

\textcolor{black}{In summary, the paper presents an easy-to-follow description of the most recent 3GPP agreements on IAB, gives  cost/performance comparisons between the IAB and fiber-connected networks, and verifies the robustness of the network to different environmental effects, which makes the paper completely different from the related literature.} 

As we demonstrate, \textcolor{black}{ along with microwave backhauling,} IAB is a cost-effective \textcolor{black}{complement} of fiber, especially in dense metropolitan areas. Moreover, independently of the cost, IAB is an appropriate tool in a number of use-cases of interest in 5G. Finally, \textcolor{black}{as we show,} while the coverage rate of the IAB  network is slightly affected by heavy \textcolor{black}{rainfall} in \textcolor{black}{suburban} areas, for a \textcolor{black}{broad} range of parameter settings and different environments, the blockage and the rain are not problematic for IAB networks, in the sense that their \textcolor{black}{impact} on the coverage probability is negligible. \textcolor{black}{High} levels of tree foliage, however, may reduce the coverage probability of the network, especially in \textcolor{black}{suburban} areas. 

The rest of the paper is organized as follows. Section \ref{3gpp} summarizes the key 3GPP discussions in Rel-16 and 17 on IAB. Section \ref{performance} describes the performance evaluation of IAB \textcolor{black}{networks}, compares \textcolor{black}{their} performance with those achieved \textcolor{black}{in} \textcolor{black}{hybrid IAB/fiber-connected} networks, and verifies the robustness of \textcolor{black}{the} IAB setup to different \textcolor{black}{weather and deployment} parameters. Finally, \textcolor{black}{conclusions and a number of interesting open \textcolor{black}{research problems that encourages researchers to contribute are} provided in Section \ref{conclusion}.
}
\section{IAB in 3GPP}\label{3gpp}
\begin{figure}
\centerline{\includegraphics[width=3.5in]{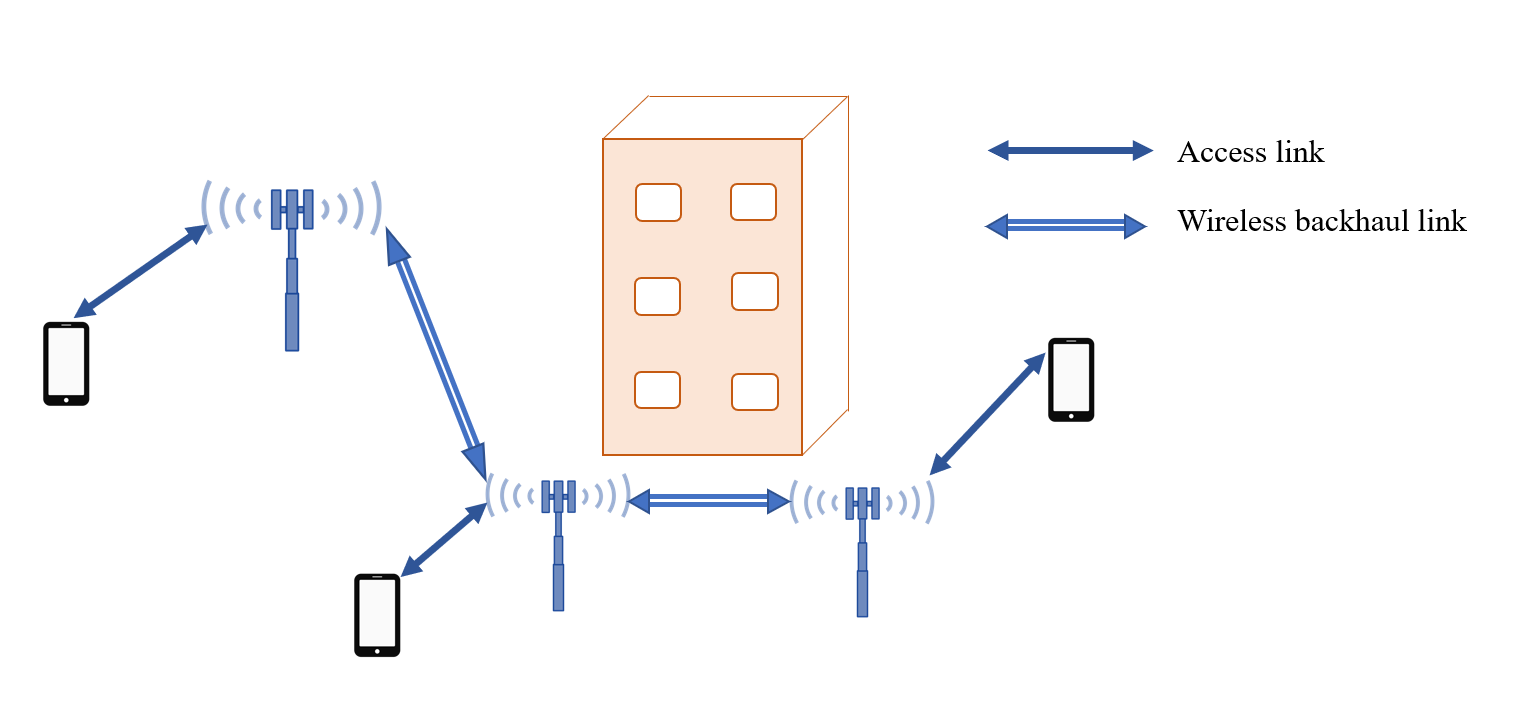}}
\caption{
Integrated access and backhaul.\label{iab3gpp1}}
\end{figure}

NR IAB was introduced in 3GPP Rel-16. It provides functionality that allows for the use of the NR radio-access technology not only for the link between BSs and devices, sometimes referred to as the access link, but also for wireless backhaul links, see Figure \ref{iab3gpp1}.

Wireless \textcolor{black}{backhauling}, that is, the use of wireless technology for backhaul links, has been used for many years. However, this has then been \textcolor{black}{mainly} based on radio technologies different from those used for the access links. \textcolor{black}{Additionally}, wireless backhaul has typically been based on proprietary, \textcolor{black}{i.e.,} non-standardized\footnote{\textcolor{black}{Some aspects of microwave backhauling are standardized, but there is significant room for proprietary solutions.}}, radio technology operating in mmWave spectrum above 10 GHz\footnote{\textcolor{black}{Traditional wireless backhaul operates also below 10 GHz, for example the longhaul links are typically at 6 GHz \cite{refi20berc}.}}
 and constrained to line-of-sight (LOS) propagation conditions.

However, \textcolor{black}{along with massive amount of available spectrum due to the move to mmWave,} there are at least two factors that now make it more relevant to consider an \textcolor{black}{IAB} solution, that is, reusing the standardized cellular technology, \textcolor{black}{normally} used by devices to access the network, also for wireless-backhaul links: 
\begin{itemize}
    \item With the emergence of 5G NR, the cellular technology is extending into the mmWave spectrum, \textcolor{black}{a} spectrum range \textcolor{black}{that} historically is used for wireless backhaul.
    \item With the emergence of small-cell deployments with BSs located, for example, on street level, there is a demand for a wireless-backhaul solution that allows for backhaul links to operate also under \textcolor{black}{non-line-of-sight (NLOS)} conditions, the kind of propagation scenarios for which the cellular radio-access technologies have been designed.
\end{itemize}
\subsection{IAB Architecture}
The IAB standard that is being specified in 3GPP Rel-16 \cite{refiab1} is based on the split architecture introduced in 3GPP Rel-15, where a base station (gNB) is split into a centralized unit (CU), which terminates the Packet Data Convergence Protocol (PDCP) and the Radio Resource Control (RRC) protocol, and a distributed unit (DU) that terminates the lower layer protocols, i.e., Radio Link Control (RLC), Medium Access Control (MAC) and the physical layer \cite{refiab2}. The motivation for the CU/DU functional split is that all time-critical functionalities, e.g.,  scheduling, fast retransmission, segmentation etc., can be realized in the DU, i.e., close to the radio and the antenna, while it is possible to centralize and resource-pool the less time-critical radio functionalities in the CU. A specified interface (F1 interface) is used to convey both the control-plane (F1-C) and user-plane (F1-U) messages between the CU and DU. The CU/DU split is transparent to the UE, i.e., it does not impact UE functionality or protocol stack. 
\begin{figure}
\centerline{\includegraphics[width=3.5in]{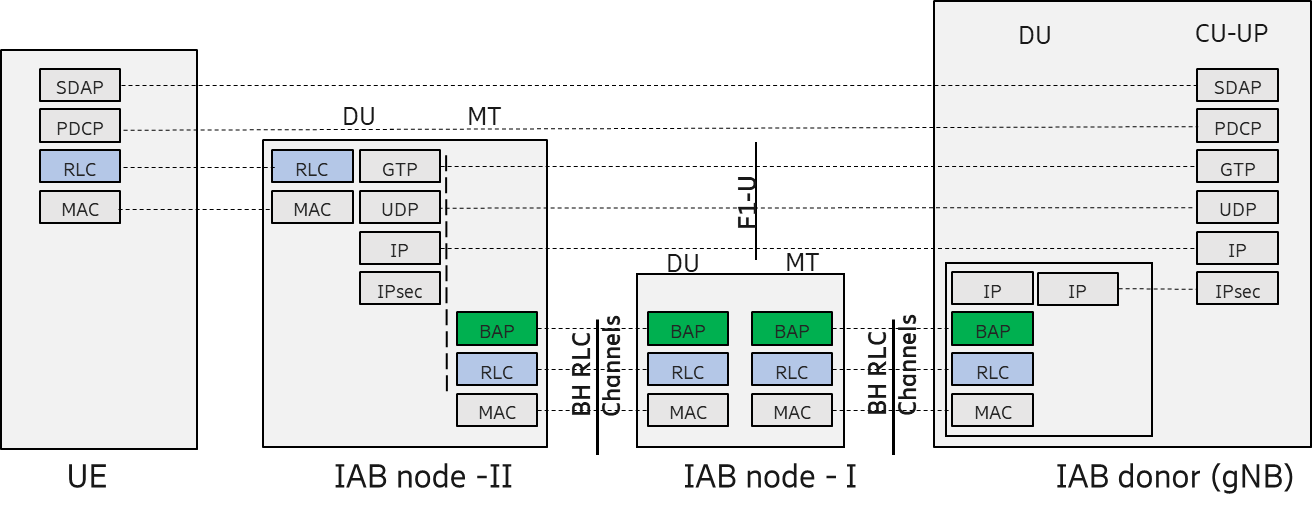}}
\centerline{\includegraphics[width=3.5in]{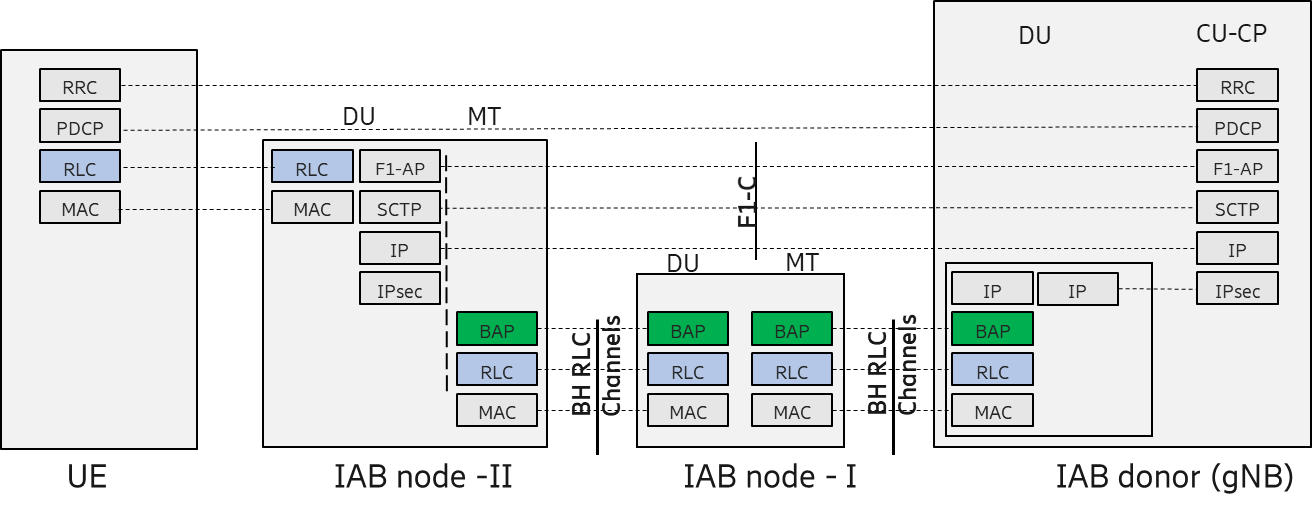}}
\caption{
User plane and control plane protocol stack of a multi-hop IAB network according to 3GPP Rel-16.\label{iab3gpp2}}
\end{figure}

Figure \ref{iab3gpp2} shows the control and user plane protocol stack of a multi-hop IAB network according to 3GPP Rel-16. The IAB donor node is the node that is connected to the rest of the network in a conventional way (e.g., fiber \textcolor{black}{or microwave}) and serves the IAB nodes and other UEs that are directly connected to it. The IAB nodes have a mobile termination (MT) part and a DU part. The MT part is used to connect to a parent DU (which could be the donor DU or the DU part of another IAB node), while the DU part of an IAB node is used to serve UEs or the MT part of child IAB nodes.

In many respects, the MT part of an IAB node behaves like a UE in the sense that it communicates with \textcolor{black}{the parent DU} very much like a UE. On the other hand, from the UE point-of-view, the DU of an IAB node appears as a normal DU. This is necessary to preserve backwards compatibility \textcolor{black}{so} that legacy \textcolor{black}{(pre Rel-16)} NR UEs \textcolor{black}{could also} access the network via an IAB node.

As in legacy CU/DU split, for the user plane, the service data adaptation protocol (SDAP) and PDCP are terminated at the UE and the user plane part of the CU (CU-UP), and the corresponding packets are transported over an F1-U interface (basically, a set of GTP tunnels for each bearer) between the CU-UP and the \textcolor{black}{DU part of the} IAB node serving the UE (known as access IAB node).  Similarly, for the control plane, the RRC and PDCP are terminated at the UE and the CU-CP, and the corresponding packets \textcolor{black}{are} transported over an F1-C interface, which is realized via a set of stream control transport protocol (SCTP) associations/streams between the CU-UP and the \textcolor{black}{DU part of the} access IAB node.  The IAB-MTs can employ all the functionalities available to UEs such as carrier aggregation and dual connectivity to multiple parent nodes. The IAB \textcolor{black}{nodes's} protocol/architecture is transparent to the UE, i.e., UEs cannot differentiate between normal gNBs and IAB nodes.

In Rel-16, only a directed acyclic graph (DAG) multi-hop topology \textcolor{black}{was supported}, i.e., no mesh-based \textcolor{black}{connectivity}. \textcolor{black}{Also, only decode-and-forward relaying was considered, where the signal is decoded in each hop and, with a successful decoding, it is re-encoded and transferred to the next hop. Compared to other relaying techniques, this gives the best E2E performance in the multi-hop setup and make it possible to scale the network to the cases with different numbers of hops,} \textcolor{black}{especially, due to its full processing capability.} The IAB nodes are interconnected with each other at layer 2 level and a hop-by-hop RLC is employed. This provides a better performance than having an \textcolor{black}{end-to-end} (E2E) RLC \textcolor{black}{between the donor and the UE} because retransmissions, if any, are required only over the affected hop, rather than between the UE and the donor, leading to faster and most efficient recovery to transmission failures. Hop-by-hop RLC also leads to \textcolor{black}{lower} buffering requirements at the end points. With regard to security, no hop-by-hop security is needed between the IAB nodes since the PDCP at the UE and CU ensure E2E encryption and integrity protection (optional for \textcolor{black}{user plane}).

\textcolor{black}{\subsubsection{On Backhaul Adaptation Protocol}}
A new protocol known as backhaul adaptation protocol (BAP) is specified that is responsible for the forwarding of packets in the intermediate hops between the donor DU and the access IAB node \cite{refiab3}.  Each IAB node is configured with a unique BAP ID by the donor node. For downlink (DL) packets, the donor DU inserts a BAP routing ID on the packets it is forwarding to the next hop, which is the BAP ID of the access IAB node serving the UE and a path identifier, in case there are several possible paths to reach the access IAB node. Similarly, for uplink (UL) packets, the access IAB node inserts the UL BAP routing ID, which is the BAP ID of the donor DU and a path identifier, in case there are several possible paths to reach the DU. Each IAB node is configured with UL and DL routing tables, which indicates to which child node (in the case of DL) or parent node (in the case of UL) the packet should be forwarded. When an access IAB node receives a packet that is destined to it, the packet will be forwarded to higher layers and processed the same way a normal DU processes incoming F1-U or F1-C packets.

In addition to forwarding packets to \textcolor{black}{a} child or parent node, the BAP protocol also performs the mapping between ingress and egress backhaul RLC channels, to ensure that the packets are treated with the proper \textcolor{black}{quality of service (QoS)} requirements. Similar to RLC channels between a DU and a UE, the backhaul RLC channels can be configured with different QoS parameters such as priority and guaranteed bit rates. For bearers that have very strict QoS requirements, a 1:1 mapping could be used, where there is a dedicated backhaul RLC channel on each hop. Otherwise, an 1:N mapping can be employed where packets belonging to several bearers could be transported/multiplexed over a given backhaul RLC channel. Similar to the routing table, the IAB nodes are configured with a mapping configuration to determine which egress backhaul RLC channel a packet should be forwarded to once the next child/parent node has been identified via the routing table.

\begin{figure}
\centerline{\includegraphics[width=3.5in]{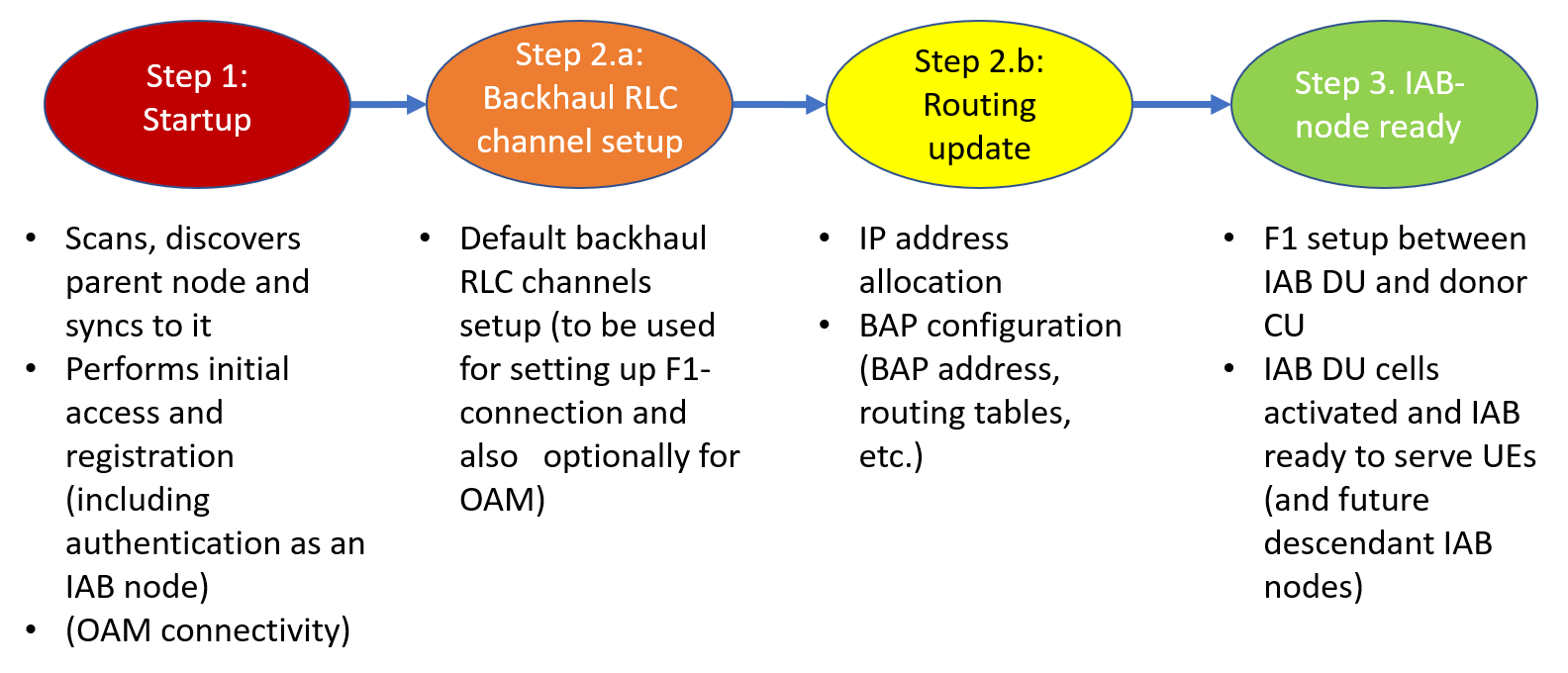}}
\caption{
Schematic diagram of the IAB integration procedure in 3GPP Rel-16. \label{iab3gpp3}}
\end{figure}

\textcolor{black}{\subsubsection{On Integration Procedure}}
Before becoming fully operational, the IAB node performs the IAB integration procedure, which is illustrated in Fig. \ref{iab3gpp3} (interested reader is referred to \cite{refiab4} for the details). In the first step (startup), the IAB node performs an RRC connection establishment\textcolor{black}{, like} a normal UE, using its MT functionality. Once the connection is set up, it indicates to the network that it is an IAB node, which the network verifies/authenticates. Connectivity to the Operation and Maintenance (OAM) part of the network could also be performed at this phase to update configurations.

In the second step, the required/default backhaul RLC channel(s) are established, to enable the bootstrapping process where the DU part of the IAB node establishes the F1 connection with the donor as well as enable OAM connectivity (if not performed during the first step).  A routing update is also made, which includes several sub-procedures such as IP address allocation for the IAB node and the (re)configuration of the BAP sub-layer at the IAB node and possibly all ancestor IAB nodes (BAP routing identifier(s) for downstream/upstream directions, routing table updates, etc.).

In the last step, the DU part of the IAB node can initiate an F1 connection request towards the donor CU, using its newly allocated IP address. After the F1 connection is set up, the IAB node can start serving UEs \textcolor{black}{like} a normal DU. Reconfigurations can be made anytime after this step, on a need basis, to update the backhaul RLC channels, routing tables, bearer mapping, etc.

\subsection{Spectrum for IAB}
\textcolor{black}{As already mentioned,} although IAB supports the full range of NR spectrum, for several reasons the mmWave spectrum is most relevant for IAB:

\begin{itemize}
    \item The potentially large amount of mmWave spectrum makes it more justifiable to use part of the spectrum resources for wireless backhaul. 
    \item Massive beamforming enabled at higher frequencies is especially beneficial for the   wireless-backhaul scenario with stationary nodes at both ends of the radio link.
\end{itemize}
\begin{figure*}[t]
\centerline{\includegraphics[width=7.5in]{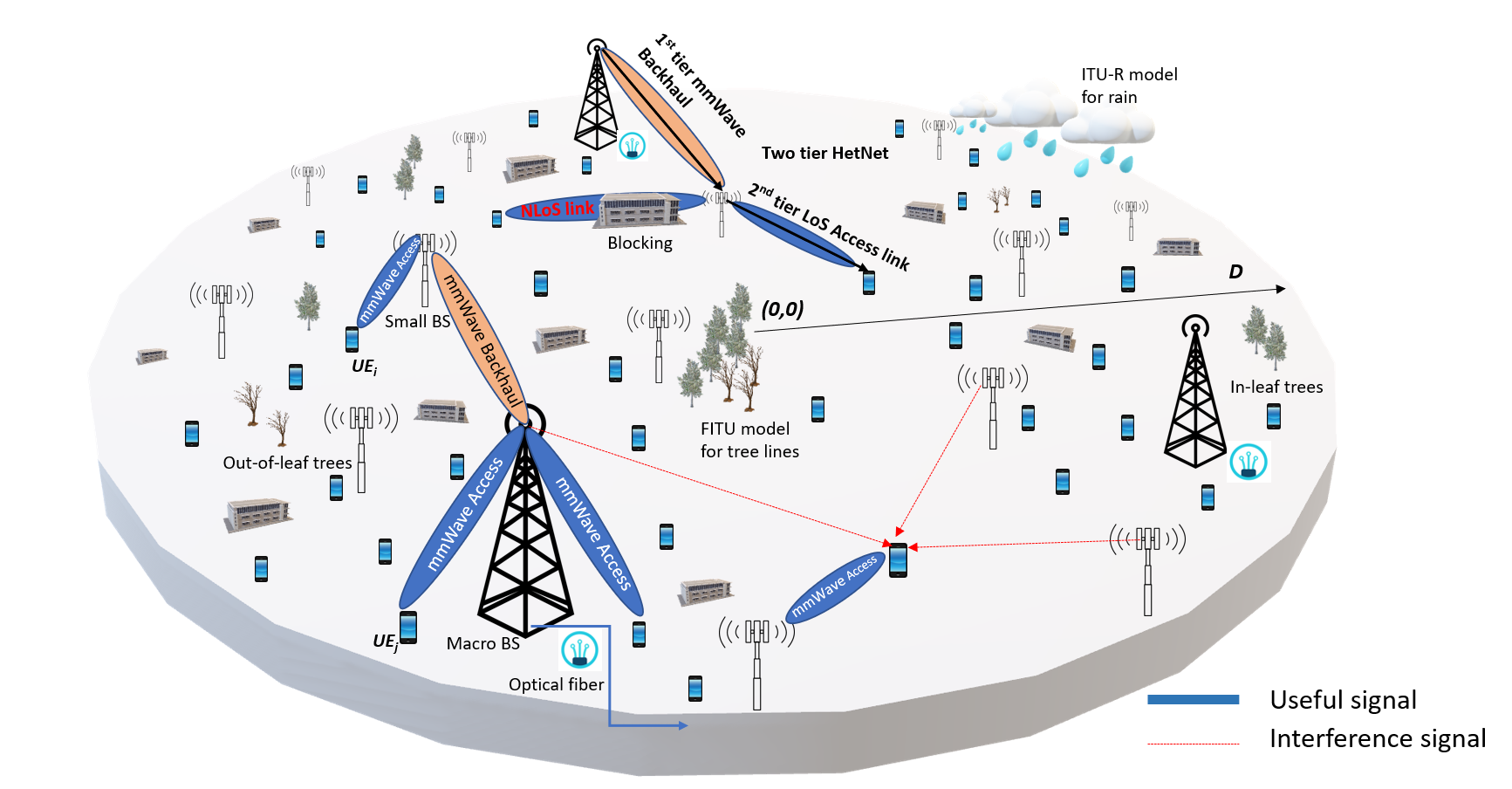}}
\caption{Schematic of the \textcolor{black}{IAB} system model.\label{systemmodelfig}}
\end{figure*}

Higher-frequency spectrum is mainly organized as unpaired spectrum. Thus, operation in unpaired spectrum has been the main focus for the 3GPP discussions on IAB. 
IAB supports both outband and inband backhauling: 
\begin{itemize}
    \item Outband backhauling\textcolor{black}{: The} wireless backhaul links operate in a different frequency band, compared to the access links.
    \item Inband backhauling\textcolor{black}{: The} wireless backhaul links operate \textcolor{black}{in} the same frequency band, as the access links.
\end{itemize}
\subsection{The IAB Radio Link}
In most respects, the \textcolor{black}{backhaul} link, between a parent-node DU and a corresponding \textcolor{black}{child IAB}-node MT operates as a conventional network-to-device link. \textcolor{black}{Consequently}, the IAB-related extensions to the NR physical, MAC, and RLC layers are relatively limited and primarily deal with the need to coordinate the IAB-node MT and DUs for the case of inband operation when simultaneous DU and MT operation is not possible.

Similar to UEs, a time-domain resource of an IAB-node MT can be configured/indicated as:
\begin{itemize}
    \item Downlink (DL): The resource will only be used by the parent node in the DL direction.
    \item Uplink (UL): The resource will only be used by the parent node in the UL direction.
    \item Flexible (F). The resource may be used in both the DL and UL directions with the instantaneous transmission direction determined by the parent-node scheduler.
\end{itemize}
Similarly, \textcolor{black}{the} time-domain resources of the DU part of an IAB node can be configured as: 
\begin{itemize}
    \item \textcolor{black}{Downlink (DL)}: The DU can only use the resource in the DL direction.
    \item Uplink (UL): The DU can only use the resource in the UL direction.
    \item Flexible (F): The DU can use the resource in both the DL and UL directions. 
\end{itemize}

In parallel to the DL/UL/F configuration, DU time-domain resources \textcolor{black}{could} be configured as hard or soft. \textcolor{black}{In case of a hard configuration, the DU of a node can use the resource without having to consider the impact on its MTs ability to transmit/receive according to its configuration and scheduling.} In practice this \textcolor{black}{means} that, if a certain DU time-domain resource is configured as hard, the parent node must assume that the IAB-node MT may not be able to receive/transmit.  Consequently, the parent node should not schedule transmissions to/from the MT in this resource.

In contrast, in case of a DU time-domain resource configured as soft, the DU can use the resource if and only if this does not impact the MTs ability to transmit/receive according to its configuration and scheduling. This means that the parent node can schedule a DL transmission to the MT in the corresponding MT resource and assume that the MT is able to receive the transmission. Similarly, the parent node can schedule MT UL transmission in the resource and assume that the MT can carry out the transmission.

The possibility to configure soft DU resources allows for more dynamic resource utilization. Take, as an example, a soft DU resource corresponding to an MT resource configured as UL. If the MT does not have a scheduling grant for that resource, the IAB node knows that the MT will not have to transmit within the resource. Consequently, the DU can dynamically use the resource, for example, for DL transmission, even if the IAB node is not capable of simultaneous DU and MT transmission.

The possibility to configure soft DU resources also gives an IAB node the chance to benefit from being able \textcolor{black}{to perform} simultaneous DU and MT operation. Whether or not a specific IAB node is capable of simultaneous DU and MT operation may depend on the IAB-node implementation and may also depend on the exact deployment scenario. Thus, an IAB node designed or deployed so that it can support simultaneous DU and MT operation can use a soft DU resource without the parent node even knowing about it. 
\begin{table*}[t]
\centering
\caption{The Definition of the Parameters.}
\setlength{\tabcolsep}{3pt}
\begin{tabular}{|p{40pt}|p{160pt}|p{40pt}|p{160pt}|}
\hline
Parameter& 
Definition&
Parameter&
Definition\\
\hline
 $\phi_{\text{M}}$ & FHPPP of the MBSs & $\phi_{\text{U}}$& FHPPP of the UEs\\$\phi_{\text{S}}$& FHPPP of the SBSs& $\lambda_{\text{U}}$& UE density\\ $\phi_{\text{B}}$&FHPPP of the blocking walls& $\theta$& Orientation of blocking wall\\$\phi_{\text{T}}$&FHPPP of the tree lines& $l_{\text{hop}}$& Average hop length\\ $\lambda_{\text{M}}$& MBSs density& $\lambda_{\text{S}}$& SBSs density\\  $H$& Homogeneous Poisson Process& $\lambda_{\text{B}}$& Blocking wall density\\ $l_{\text{B}}$& Blocking wall length& $\rho$&Service coverage probability\\$A$&Circular disk&$D$&Radius of the disk\\$P_{\text{t}}$&Transmission power&$P_{\text{r}}$&Received power\\$h$&Fading coeficient&$G$&Antenna gain \\${L}_{(1\text{m})}$&Reference path loss at 1 meter distance&$L_{}$&Propagation path loss\\$x$&Location of the node&$r$&Propagation distance  between  the  nodes\\$\alpha$&Path  loss  exponent&$N$&Number of UEs connected\\$f_{\text{c}}$&Carrier frequency& $\varphi$&Angle between the BS and UE\\$\theta_\text{{HPBW}}$&Half power beamwidth of the antenna&$G_0$&Maximum gain of directional antenna\\$g(\varphi)$&Side lobe gain&$R_{\text{th}}$& Minimum data rate threshold\\$x_{\text{c}}$&Associated cell&$R$&Rain intensity\\$x_{\text{u}}$&Connected UE at the location & $i,j$ & Node index\\$F_\text{T}$&Tree foliage&$\gamma_{\text{R}}$&Rainloss\\$d$&Vegetation depth&$W$&Bandwidth of the DL \\$\mu$&Percentage of bandwidth resources on backhaul&$l_{\text{T}}$&Tree line length\\$\lambda_{\text{T}}$&Tree blocking density & $v$ & SBS antenna height

\\

\hline

\end{tabular}
\label{tabpar}
\end{table*}

These situations, when an IAB node, by itself, can conclude that it can use a soft DU resource has, in the 3GPP discussions, been referred to as implicit indication of availability of soft DU resources. The parent node can also provide an explicit indication of availability of a soft DU resource by means of layer-1 signaling.

Finally, it should be noted that, along with resource multiplexing which has been the main topic of discussions in RAN1, the over-the-air (OTA) timing alignment, the random access channel (RACH) as well as the extensions of SSBs for inter-IAB-node discovery and measurements have been discussed in \textcolor{black}{3GPP}. However, due to space limits, we do not cover these topics, and the interested reader can find the final agreements in \cite{refiab5}. \textcolor{black}{Moreover, while we concentrated \textcolor{black}{mostly} on RAN1 and RAN2 discussions, the main discussions/agreements in RAN3 and RAN4 can be found in \cite{refiab5erc1}-\cite{refiab5erc2} and \cite{refiab5erc3}, respectively (also see \cite{refi6}).}
\subsection{IAB in Rel-17}
The physical-layer part of the IAB Rel-16 specifications was finalized at the end of 2019 and the remaining parts (higher-layer protocols and architecture) \textcolor{black}{are expected to be finalized} in June 2020. Further enhancements to IAB will then be carried out within 3GPP Rel-17, with expected start in August 2020 \cite{refiab6}. The Rel-17 work aims to improve on various aspects such as robustness, degree of load-balancing, spectral efficiency, multi-hop latency and end-to-end performance. More specifically, the following is planned to be covered: 
\begin{itemize}
    \item Enhancements to the resource multiplexing between child and parent links of an IAB node, including:
    \begin{itemize}
        \item Enhanced support of simultaneous operation (transmission and/or reception) of IAB-node’s child and parent links, including enhancements such as new DU/MT timing relations,  DL/UL power control and cross link interference mitigation.
        \item Support for dual-connectivity scenarios for topology redundancy for improved robustness and load balancing.
    \end{itemize}
    \item Enhancements in scheduling, flow and congestion control to improve end-to-end performance, fairness, and spectral efficiency. 
    \item Introduction of efficient inter-donor IAB-node migration, increasing the robustness of IAB networks allowing for more refined load-balancing \textcolor{black}{and topology management.}
    \item Reduction of service interruption time caused by IAB-node migration and backhaul RLF recovery improves network performance, allows network deployments to undergo more frequent topology changes, and provides stable backhaul performance.
\end{itemize}

\textcolor{black}{Finally, it should be mentioned that in 3GPP RAN4 a number of simulations have been performed to evaluate the feasibility/efficiency of IAB networks, e.g., \cite{refiab6erc}. There, it has been mainly concentrated on defining RF requirements for both backhaul and access links of an IAB-node including requirements for their co-existence, and evaluate the performance in different possible scheduling scenarios of the the DU and MT.} \textcolor{black}{In Section \ref{performance}, we} mainly concentrate on the comparison between the performance of IAB and fiber networks as well as studying the robustness of IAB networks to different environmental effects \textcolor{black}{using a novel stochastic geometry modeling for mmWave networks and 3D maps topology information.} Such results provide insights about if the IAB performance expectations will be met in urban and suburban areas.

\section{Performance Evaluation}\label{performance}

This section studies the service coverage rate \textcolor{black}{of IAB networks with various parameterizations, and compares the performance} with those achieved by (partially) fiber-connected networks. First, we present the system model, including the channel model, the considered UE association rule as well as the rain, the blockage and the tree foliage models, which are followed by the simulation results.

\subsection{System Model}\label{systemmodel}
\textcolor{black}{As shown in Fig. \ref{systemmodelfig}., consider an outdoor two tier heterogeneous network (HetNet), i.e., a two-hop IAB network, with multiple MBSs (M: macro), SBSs (S: small) and UEs. This is motivated by different evaluations, e.g., \cite{refi17,refi18,refi19,refi20}, where, although the standardization does not limit the number of hops, increasing the number of hops may lead to backhaul traffic aggregation.} In an IAB deployment, both the MBSs and the SBSs use wireless connections for both access and backhaul. Also, only the MBSs are fiber-connected while the SBSs receive data from the MBSs wirelessly \textcolor{black}{by using IAB}. That is, following the 3GPP definitions (see Section \ref{3gpp}), the MBSs and the SBSs can be considered as the donor and the child IABs, respectively. Therefore, throughout the section, we may use the terminologies MBS/SBS and donor IAB/IAB  interchangeably. \textcolor{black}{Considering an \textcolor{black}{inband} operation, the} bandwidth is shared among access and backhaul links of the IAB nodes such that the network service coverage rate is maximized. \textcolor{black}{For simplicity,} the MBSs and the SBSs are assumed to have constant power over the spectrum of the system and are all active throughout the analysis\footnote{\textcolor{black}{Developing adaptive power allocation schemes for IAB networks is an interesting open research topic.}}.

\subsubsection{Spatial Model}
Table I summarizes the parameters used in the analysis. We model the IAB network by \textcolor{black}{an} FHPPP, e.g., \cite{refnewb}, \cite{ref1}, which suits well to model a random number of nodes in a finite region. Particularly, FHPPPs $\phi_{\text{M}}$ and $\phi_{\text{S}}$ with densities $\lambda_{\text{M}}$ and $\lambda_{\text{S}}$, respectively, are used to model the spatial distributions of the MBSs and the SBSs, respectively.

The MBSs' FHPPP is given by $\phi_{\text{M}}=H\cap A$, where $H$ with density $\lambda_{\text{M}}$ is an HPPP (H: homogeneous) and $A\subset\mathbb{R}^2$ is a finite region. For simplicity and without loss of generality, we let $A$ be a circular disk with radius $D$. However, the study is generic and can be applied on arbitrary regions $A$. The SBSs and the UEs are also located within the same $A$ in accordance with two other FHPPPs $\phi_{\text{S}}$ and $\phi_{\text{U}}$ having densities $\lambda_{\text{S}}$ and $\lambda_{\text{U}}$, respectively, which are all mutually independant.

We study the system performance for two blocking conditions. First, we use the well-known germ grain model \cite[Chapter 14]{ref2}, which provides accurate results compared to stochastic models that assume the blocking in different links to be independant. Moreover, the  germ grain model fits well for environments with large obstacles as it takes the obstacles induced blocking correlation \textcolor{black}{into} account. The model is \textcolor{black}{an} FHPPP, i.e., the blockages are distributed according to the FHPPP $\phi_{\text{B}}$ distributed in the same area $A$ with density $\lambda_{\text{B}}$. This is a 2D model where all blockings are assumed to be walls of length $l_{\text{B}}$ and orientation $\theta$, which is an independantly and identically distributed (IID) uniform random variable in [0, 2$\pi$].
The walls are distributed in random locations uniformly as of the FHPPP. 

With the 2D channel model, the elevation of the blocking and the BSs or the terrain information of the land are not taken into
account. For this reason, in Subsection \ref{3d}, we demonstrate the system performance for an example 3D use-case. Particularly, we distribute the same spatial arrangement of the MBSs, the SBSs and the UEs with their respective nodes heights on top of map data with real world blocking terrain using OpenStreetMap 3D environment. That is, while different MBS and SBS nodes are distributed randomly based on their corresponding FHPPPs, they are placed, on different heights, and the blockages are determined based on the map information. This enables us to evaluate the effect of the nodes and blocking heights on the service coverage probability.

\subsubsection{Channel Model}\label{AA}

We consider an inband communication setup, where both the access and backhaul links operate in \textcolor{black}{the same} mmWave spectrum \textcolor{black}{band}. Following the state-of-the-art mmWave channel model, e.g., \cite{ref1}, the received power at each node can be expressed as
\begin{equation}
    P_{\text{r}}=P_{\text{t}}h_{\text{t,r}}G_{\text{t,r}}{L}_{(1\text{m})}L_{\text{t,r}}\left|\left|x_{\text{t}}-x_{\text{r}}\right|\right|^{-1}F_{\text{t,r}}\gamma_{\text{{t,r}}}.
\end{equation}
Here, $P_{\text{t}}$ denotes the transmit power in each link, and $h_{t,r}$ represents the independant small-scale fading for each link. The small-scale fading is modelled as a normalized Rayleigh random variable in our analysis. Then, $G_{t,r}$ represents the combined antenna gain of the transmitter and the receiver of the link, \textcolor{black}{$L_\text{t,r}$ which is a function of the distance between $x_\text{t}$ and $x_\text{r}$, denotes the path loss due to propagation,} and ${L}_{(1m)}$ is the reference path loss at 1 meter distance. The tree foliage loss is denoted by $F_{\text{t,r}}$ while $\gamma_{\text{{t,r}}}$ represents the rain loss between the transmitter and the receiver of the link in linear scale. The total path loss, in dB, is characterized according to the 5GCM UMa close-in model described in \cite{ref3}. The path loss is given by

\begin{equation}
\textcolor{black}{\kappa}=32.4+10\log_{10}(r)^\alpha+20\log_{10}(f_\text{c}),
\end{equation}
where $f_c$ is the carrier frequency, $r$ is the propagation distance between the nodes, and $\alpha$ is the path loss exponent. 
Depending on the blockage, LOS and NLOS links are affected by different path loss exponents. The propagation loss of the path loss model is given by 
\begin{equation}L_{\text{t,r}} = \begin{cases} r^{\alpha _{\text{{L}}}},&\text {if LoS,} \\ r^{\alpha_{{{\text{N}}}}},&\text {if NLoS,}{} \end{cases}
\end{equation}
where  $\alpha_\text{L}$ and $\alpha_\text{N}$ denote path loss exponents for \textcolor{black}{the} LOS and NLOS \textcolor{black}{scenarios}, respectively. In 5G, large antenna arrays with directional beamforming are used to mitigate the propagation losses. We model the \textcolor{black}{beam pattern} as a sectored-pattern antenna array and thus the  antenna gain between two nodes can be expressed by
\textcolor{black}{
\begin{equation}G_{i,j }(\varphi) = \begin{cases} G_0&\frac{-\theta_{\text{HPBW}}}2\leq\varphi\leq\frac{\theta_{\text{HPBW}}}2 \\ g(\varphi)&\text{otherwise.} \end{cases}
\end{equation}}
Here, $i$, $j$ are the indices of the considered transmit and receive nodes, and $\varphi$ is the angle between them in the considered link. Also, $\theta_\text{{HPBW}}$ is the half power beamwidth of the antenna, and $G_0$ is the directional antenna's maximum gain while $g(\varphi)$ is the side lobe gain. \textcolor{black}{Also, we let the UE antenna gain to be 0 dB. This is in harmony with, e.g., \cite{refi17}, \cite{refi19}, \cite{paramc}, and because the UE has an omni-directional radiation pattern.} For discussions on how the antenna gain is affected by the antenna array properties, see, e.g., \cite{ref1}.

We assume that we have high beamforming capability in the IAB-IAB backhaul links. Consequently, we ignore the interference in  the backhaul links and assume them to be noise-limited. Also, the inter-UE interferences are neglected due to the low power of the devices and with the assumption of sufficient isolation  \cite{refmikpaper}. On the other hand, as illustrated in Fig. \ref{systemmodelfig}, the interference model focuses on the aggregated interference on the access links, due to the neighbouring interferers, which for UE ${u}$ is given by
\begin{equation} I_{{\rm {{u}}}}= \sum \limits _{ {\mathbf{i,j}\in \phi _{i,j}\setminus \{{\mathbf{x}}_{c}\}}}{P_{j}}  h_{i,j} G_{i,j}{L}_{(1\text{m})} L_{{x_i,x_j}}\|{\mathbf{x_\textit{i}-x_\textit{j}}}\|^{-1}.
\label{inter}
\end{equation}
Here, $i$ and $j$ represents all BSs except for the associated cell $x_\text{c}$ which can either be \textcolor{black}{an MBS or an SBS}.

\subsubsection{Rain and Tree Foliage Model}\label{AA1}

With the need of understanding the performance of IAB \textcolor{black}{networks} in rainy conditions, we use the ITU-R Rec 8.38-3 rain model \cite{ref4} to entail the rain effect on the links. This is an appropriate model used to methodically determine the amount of rain attenuation on radio links. The model is widely used in all regions of the world, for the frequency range from  1 GHz to 1000 GHz with no rain rate obligation. The model describes the rain loss as
\begin{equation}
    \gamma_{\text{R}}=kR^\beta,
\end{equation}
where $\gamma_\text{R}$ is the rain loss in dB/km, and $R$ is the rain intensity in mm/hr. Moreover, $k$ and $\beta$ are coefficients that are precalculated  depending on the carrier frequency. Table II shows the coefficients for horizontal and vertical losses at rainy conditions in 28 GHz on which we concentrate in the simulations.

\begin{table}[btp]\begin{center}
\caption{Coefficients for ITU-R model. Here, $\beta_h$, $k_h$ are the horizontal polarization coefficients and $\beta_v$, $k_v$ denote the vertical polarization coefficients \cite{ref4}. }
\setlength{\tabcolsep}{3pt}
\begin{tabular}[btp]{|p{75pt}|p{25pt}|p{25pt}|p{25pt}|p{25pt}|}
\hline
Frequency (GHz)& 
$\beta_h$& $\beta_v$& $k_h$& $k_v$ \\
\hline
$28 $& 
0.9679& 0.9277& 0.2051& 0.1964
\\
\hline

\end{tabular}\end{center}
\label{itutab}
\end{table}

Finally, FHPPP $\phi_{\text{T}}$ with density $\lambda_{\text{T}}$ is used to spatially distribute the tree lines of length $l_\text{T}$ \cite{reftree}. We use the Fitted International Telecommunication Union-Radio (FITU-R) tree foliage model \cite[Chapter 7]{ref5} to model the effect of  the trees on the received signal power. This is an appropriate model for the cases with frequency dependancy and with non-uniform vegetation. The model \textcolor{black}{is suitable} for the mmWave frequencies from 10 to 40 GHz and has been derived by further developing the ITU-R vegetation model. In this way, considering two, namely, \textcolor{black}{\textit{in-leaf} and \textit{out-of-leaf}}, vegetation states, the tree foliage  in dB is obtained by
\begin{equation}
   F_\text{T}=\left\{\begin{array}{l}0.39f_{\text{c}}^{0.39}d^{0.25},\;\text{in-leaf}\\0.37f_{\text{c}}^{0.18}d^{0.59},\;\text{out-of-leaf,}\;\end{array}\right.
   \label{eqvege}
\end{equation}
where $f_{\text{c}}$ is the carrier frequency expressed in MHz and $d$ is the vegetation depth in \textcolor{black}{meter}.

\subsection{Association and Allocation Strategy}

In our setup, the UE can be served by either \textcolor{black}{an MBS or an SBS} following open access strategy and based on the maximum average received
power rule. Also, in harmony with 3GPP, we do not take joint transmission into account, i.e., each UE can be connected to only one MBS or SBS. In this way, the association rule for UE $u$ suffices 
\begin{equation}
    \sum_{\forall j}u_{j}=1,\;\forall_u\in U,u_i\cdot u_j=0, \forall j\ne i,\;
\end{equation}
where $u_{j}\in \{0,1\}$ is a binary variable indicating the association with 1 and 0 denoting the unassociated cell.
For the access links of the UEs, we have
\textcolor{black}{
\begin{equation}\begin{aligned} {u_j}  = \begin{cases} 1 & \text {if}~P_{i} G_{z,x}h_{z,x}{L}_{(1m)} L_{{\rm {z,x}}}(\|{\mathbf{z}}-{\mathbf{x}}\|)^{-1} \\&\qquad \quad ~~\geq P_{j}G_{j}h_{z,y}{L}_{(1m)} L_{{\rm {z,y}}}(\|{\mathbf{z}}-{\mathbf{y}}\|)^{-1}, \\&\qquad \qquad \quad \forall ~{\mathbf{y}}\in \phi _{j}|{\mathbf{x}}\in \phi _{i}, i,j\in \{{\mathrm{ m}},{\mathrm{ s}}\} \\ 0,& \text {otherwise.} \end{cases}\end{aligned} \label{eq:9}
\end{equation}}
As in \eqref{eq:9} for each UE ${u}$, the association binary variable $u_j$ becomes 1 for the cell giving the maximum received power at the UE, while for all other cells it is 0 since the UE can only be connected to one IAB node. 

Because the IAB nodes, i.e., both the MBSs and the SBSs, are equipped with large antenna arrays and can beamform towards the required direction, the antenna gain over the backhaul links can be assumed to be the same, and backhaul link association can be well determined  based on the minimum path loss rule, i.e., by
\begin{equation}\begin{aligned}
{x_{b,m}} = \begin{cases} 1 & \text {if}~L_{{\rm {b}}_{\mathrm{ m}}}(\|{\mathbf{z}}-{\mathbf{x}}\|)^{-1}\!\geq \! L_{{\rm {b}}_{\mathrm{ m}}}(\|{\mathbf{z}}\!-\!{\mathbf{y}}\|)^{-1}, \\ &\qquad \qquad \qquad \qquad \qquad \forall ~{\mathbf{y}}\in \phi _{\mathrm{ m}}|{\mathbf{x}}\in \phi _{\mathrm{ m}}, \\ 0,& \text {otherwise} \end{cases} \end{aligned} 
\end{equation}

\textcolor{black}{ (For the effect of interference in the backhaul links, see Fig. \ref{hybrid}).} For resource allocation, on the other hand, the mmWave spectrum available is partitioned \textcolor{black}{into} the access and backhaul links such that 
\begin{equation}\begin{aligned}
    \begin{cases}
    W_\text{Backhaul}=\mu W,\\
    W_\text{Access}=(1-\mu)W,
\end{cases}
\end{aligned}
  \label{eq:11}  
\end{equation}


 with $\mu\in [0,1]$ being the percentage of bandwidth resources on backhauling. Also, $W_\text{backhaul}$ and $W_\text{access}$ denote the backhaul and the access bandwidths, respectively, while total bandwidth is $W$. The bandwidth allocated for each SBS, i.e., child IAB, by the fiber-connected MBS, i.e., IAB donor, is proportional to its load and the number of UEs in the access link. The resource allocation is determined based on the instantaneous load in which each SBS informs its current load to the associated MBS each time.  Thus, the backhaul-related bandwidth for the $j$-th IAB node is given by  
\textcolor{black}{ 
\begin{equation}
{W_{\text{{backhaul}}{,j}}}=\frac{\mu WN_j}{{\displaystyle\sum_{\forall\;j}}N_j},{     \forall j},
\end{equation}}
\textcolor{black}{where $N_j$ denotes the number of UEs connected to the $j$-th IAB node. \textcolor{black}{Therefore, the bandwidth allocated to the $j$-th IAB node is proportional to the ratio between its load, and the total load of its connected IAB donor.} Meanwhile, the access spectrum is equally shared among the connected UEs at the IAB node according to
\begin{equation}
{W_{\text{{access}},u}}=\frac{(1-\mu)W}{{\displaystyle\sum_{\forall\;u}}N_{j,u}}, \forall u,
\end{equation}
where  $u$ represents the UEs, and $j$ represents each IAB node. Also, $N_{j,u}$ denotes the users connected to the $j$-th IAB node of which UE $u$ is connected.} The signal-to-interference-plus-noise ratio (SINR) values are obtained in accordance with \eqref{inter} by \begin{align}
\text{SINR}=P_{\text{r}}/(I_{{u}}+N_0),
\end{align}
where $N_0$ is the noise power. Then, considering sufficiently long codewords, which is an acceptable assumption in IAB networks, the rates experienced by the UEs in access links can be expressed by 
\begin{equation}
 {R_{{u}}} =   \begin{cases}\frac{(1-\mu)W}{{}N_m}\log(1+\text{SINR}(x_{u})), ~\text { if }{\mathbf{x}}_\text{c}\in \phi _{\mathrm{ m}},\\ \min \bigg (\frac{(1-\mu)W}{{\displaystyle\sum_{\forall\;u}}N_{j,u}}\log(1+\text{SINR}(x_{u})), \\ \qquad  \frac{\mu WN_j}{{\displaystyle\sum_{\forall\;j}}N_j}\log(1+\text{SINR}(x_\text{b}))\bigg ), \text {if }{\mathbf{x}}_\text{c}\in \phi _{\mathrm{ s}} \end{cases}
 \label{eq:14}
\end{equation}
and the backhaul rate is given by

\begin{equation}
{R_{\text{b}}}=\frac{\mu WN_j}{{\displaystyle\sum_{\forall\;j}}N_j}\log(1+\text{SINR}(x_{\text{b}})).
\end{equation}
\textcolor{black}{Here, $m$ represents the associated MBS and $s$ denotes the SBS. Based on the association cell, there are two cases for the rate of the connected UEs, $x_u$ at the location. First, is the case in which the UEs are associated to the MBSs, as denoted by $x_{\text{c}} \in \phi _{\mathrm{ m}}$ in \eqref{eq:14}.} Since the MBSs, i.e., IAB donor nodes, have fiber backhaul connection, the rate will depend on the access bandwidth available at the UE. In the second case, the UEs are connected to the SBSs, as denoted by $x_{\text{c}} \in \phi _{\mathrm{s}}$ in \eqref{eq:14}. Here, the SBSs have shared backhaul bandwidth from the IAB donor nodes  i.e., MBSs, and thus the UEs data rates depend on the backhaul rate of the connected SBS as well. Thus, in this case the UE is bound\textcolor{black}{ed} to get the minimum between backhaul and access rate.

\subsection{Simulation Results and Discussions}

\begin{table}
\caption{Simulation Parameters.}
\setlength{\tabcolsep}{3pt}
\begin{tabular}{|p{95pt}|p{150pt}|}
\hline
Parameters& 
Value \\
\hline
 Carrier frequency&28 GHz\\Minimum data rate threshold& 100 Mbps \\Bandwidth& 1 GHz\\ IAB node and UE density& \{MBS, SBS, UE\} = (8, 100, 500) /$\text{km}^2$\\ Blocking& \{Density, Length\} = (500 /$\text{km}^2$, 5 {\text{ m}})\\Path loss exponents& \{LoS, NLoS\} = (2, 3)\\ Main lobe antenna gains& \{MBS, SBS, UE\} = (24, 24, 0) dBi\\ Side lobe antenna gains& \{MBS, SBS, UE\} = (-2, -2, 0) dBi\\ Half power beamwidth& \{azimuthal, elevation\} = {$\text{(60, 25)}$}\\ Noise power& 5 dB\\In-leaf percentage& 20\%\\ Tree depth& 5 m\\ Antenna heights& \{MBS, SBS, UE\} = (25,10,1) m

\\

\hline

\end{tabular}
\label{tab1}
\end{table}
The simulation results are divided into three parts in which 1) we compare IAB, \textcolor{black}{hybrid IAB/fiber-connected,} and fiber-connected networks, 2) verify the robustness of IAB networks, and 3) study the system performance in an example of 3D network deployment. \textcolor{black}{Note that the 2D model is considered mainly to limit the simulation complexity. However, for different cases, the same qualitative conclusions as those presented in the 2D model hold in the 3D model as well.} The general system parameters are summarized in Table \ref{tab1} and, for each figure, the specific parameters are given in the figure caption. The network is deployed in a disk of radius of $D=1$ km, where the rain occurrence, the blockage, and the vegetation distributions are also probable according to the statistical models described in Section \ref{systemmodel}. In all figures, except for Figs. \ref{rain} and \ref{vegetation} which study the system performance in both urban and \textcolor{black}{suburban} areas, we concentrate on dense areas as the most important point of interest in IAB networks. \textcolor{black}{We assume that we have high beamforming capability in the IAB-IAB backhaul links. Consequently, in all figures except for Fig. \ref{hybrid}, we ignore the interference in the backhaul links, in harmony with, e.g., \cite{refi17,refi18,refi19}. In Fig. \ref{hybrid}, however, we verify this assumption, and study the system performance in the cases where the interference is not ignored in the IAB-IAB links (More insights on mmWave interference in cellular networks are discussed in, e.g., \cite{refin1ch,refin2ch}.)}

Our metric of interest is the service coverage probability \cite{ref7}, defined as the fraction of the UEs which have instantaneous UE data rates higher than or equal to a threshold $R_\text{th}$. That is, using \eqref{eq:14}, the service coverage probability is given by
\begin{align}
\rho= \Pr(R_\text{U} \ge R_\text{th}).
\end{align} 
\subsubsection{IAB versus Fiber}

\begin{figure}
\centerline{\includegraphics[width=3.5in]{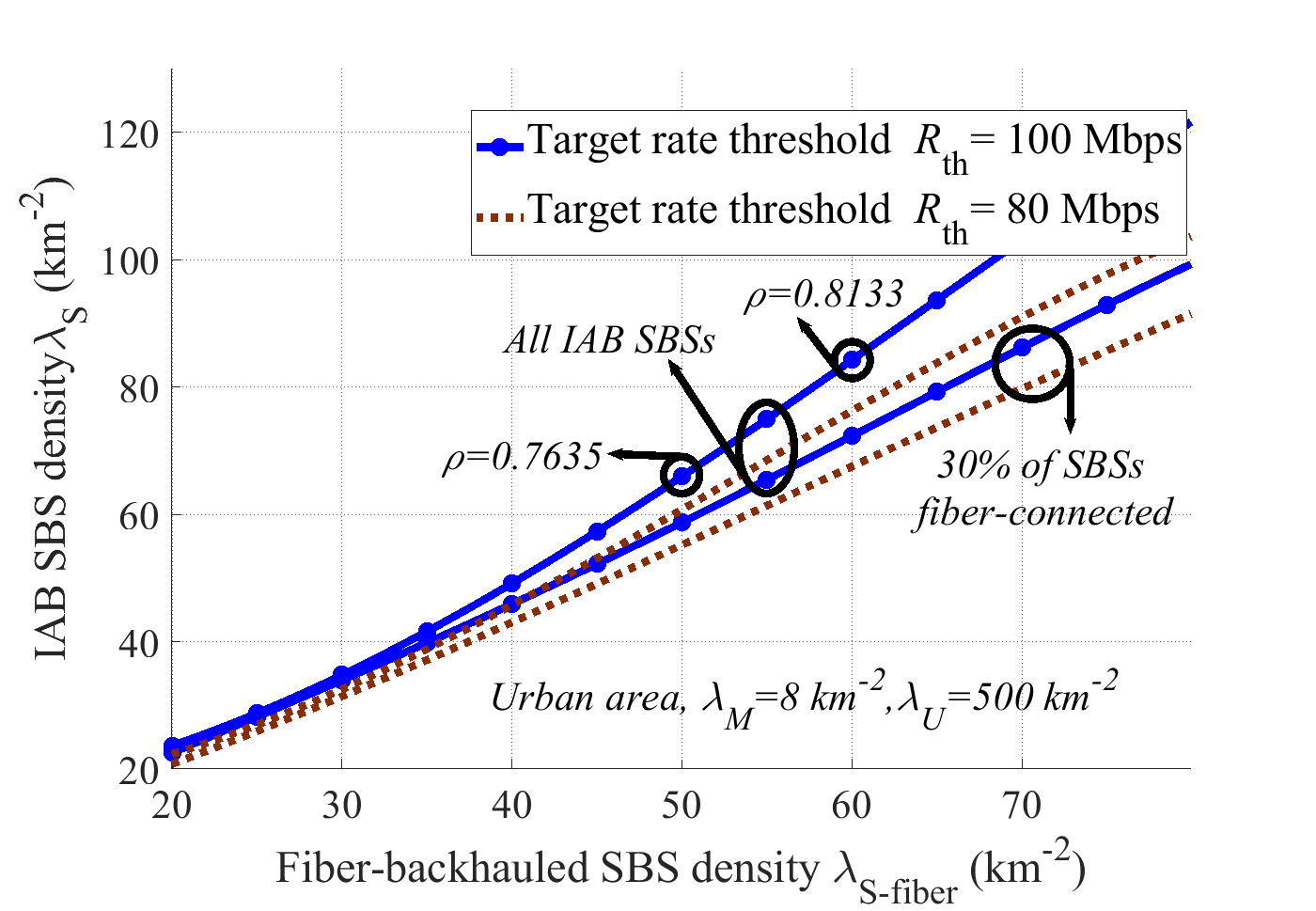}}
\caption{
 Density of the IAB nodes sufficing the performance of fiber-backhauled network, in terms of service coverage probability. The parameters are set to $\lambda_{\text{B}}=500$ $\text{km}^{-2}$, no rain, $R_{\text{th}}$ = 100 Mbps, and ${P_{\text{MBS}}, P_{\text{SBS}}, P_{\text{UE}}} = (40, 24, 0)$ dBm.\label{iabeq}}
\end{figure}

\begin{figure}
\centerline{\includegraphics[width=3.5in]{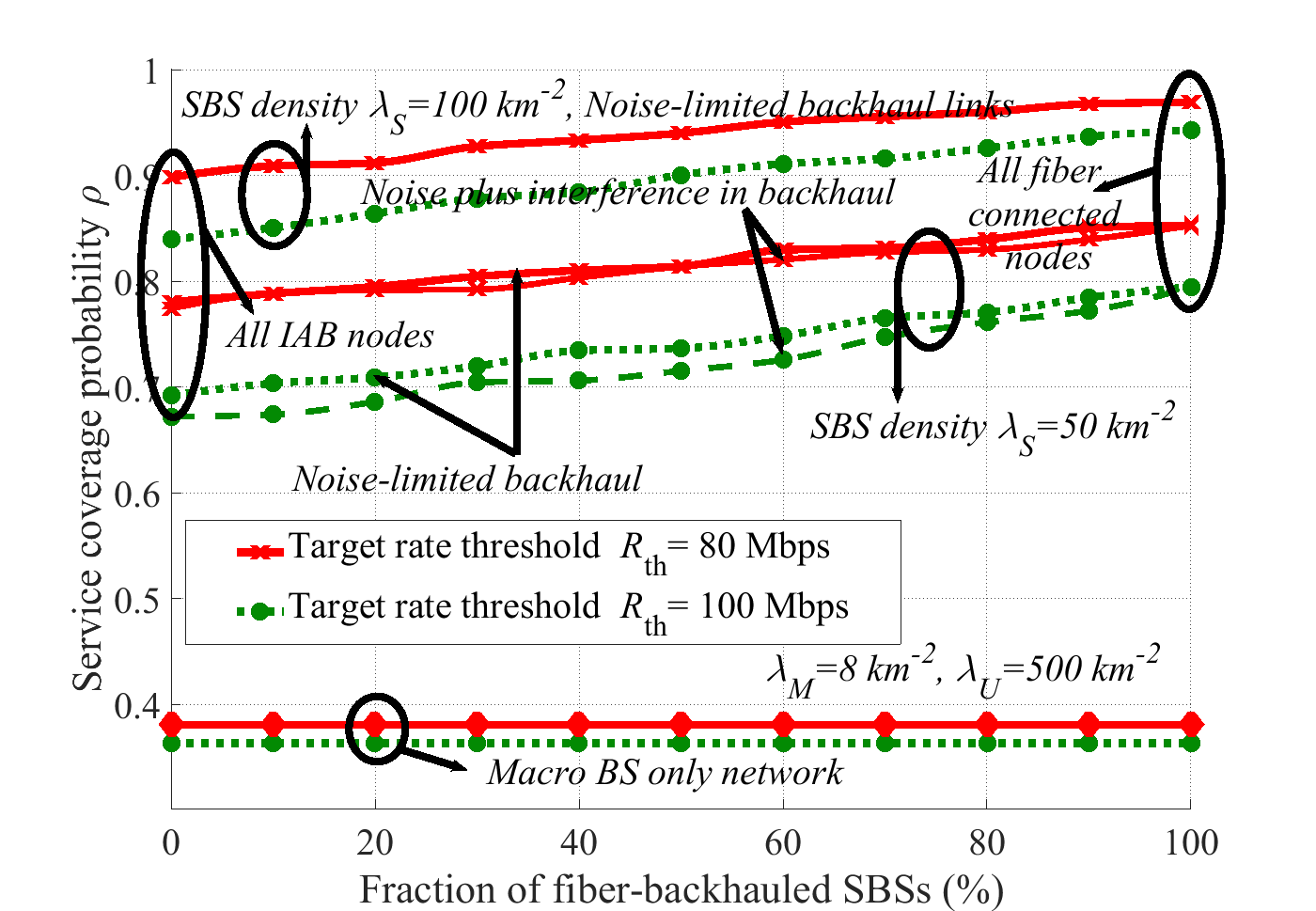}}
\caption{Service coverage probability as a function of the percentage of the fiber-backhauled SBSs for a dense network with $\lambda_{\text{B}}$ = 500$\text{ km}^{-2}$, no rain and ${P_{\text{MBS}}, P_{\text{SBS}}, P_{\text{UE}}} = (40, 24, 0)$ dBm. \label{hybrid}}
\end{figure}

In Figs. \ref{iabeq}-\ref{hybrid}, we compare the coverage probability of the IAB networks with those obtained by the cases having \textcolor{black}{(a fraction of)} fiber-connected SBSs, as well as the cases with no SBS. \textcolor{black}{Also, Fig. 6 verifies the effect of the interference in the backhaul links on the system performance.} In these figures, different parameters, e.g., bandwidth allocation between the access and backhaul, have been optimized to maximize the coverage probability in each case. Note that, in practice and depending on the network topology, a number of SBSs may also be fiber-connected. For this reason, in Figs. \ref{iabeq}-\ref{hybrid}, we also consider the cases with a fraction of SBSs having fiber connections. In such cases, we assume the fiber-connected SBSs to be randomly distributed, and adapt the association and allocation rules as well as the achievable rates, correspondingly.

Figure \ref{iabeq} demonstrates the required number of IAB nodes to guarantee the same coverage probability as in the cases with \textcolor{black}{hybrid IAB/fiber-connected} SBSs. Then, Fig. \ref{hybrid} shows the network service coverage rate as a function of the fraction of fiber-connected SBSs, and compares the system performance with the cases having no SBS.

As demonstrated in Figs. \ref{iabeq}-\ref{hybrid}, for a broad range of parameter settings, the same performance as in the fully fiber-connected networks can be achieved by the IAB network, with relatively small increment in the number of IAB nodes. As an example, consider the parameter settings of Fig. \ref{iabeq} and the UEs’ target rate 100 Mbps. \textcolor{black}{Then, a fully fiber-connected network with SBSs densities $50$ and $60$ $\text{km}^{-2}$ corresponds, in terms of coverage probability, to an IAB network having densities $\lambda_\text{S}=65$ $\text{km}^{-2}$ and 85 $\text{km}^{-2}$, respectively, leading to coverage probabilities $0.76$ and $0.81$.} Interestingly, with a $30\%$ of SBSs having fiber connections, which is practically reasonable, these numbers are reduced to $\lambda_\text{S}=70$ and $85$ $\text{km}^{-2}$, i.e., only $16\%$ and $21\%$ increase in the required number of SBSs. Then, as the network density increases, the effect of the UEs target rate as well as the relative performance gap of the IAB and fiber-connected networks decrease (Fig. \ref{hybrid}). \textcolor{black}{Moreover, in harmony with intuitions and motivated by the high beamforming capability of the IAB nodes, the effect of the interference in the backhaul links is negligible, and the IAB-IAB links can be well assumed to be noise-limited (Fig. \ref{hybrid}).}

Here, it should be noted that our results, based on the FHPPP and random node drop, give a pessimistic performance of IAB networks. In practice, the network topology will be fairly well-planned, further reducing the gap between the performance of IAB and fiber-connected networks. \textcolor{black}{Also, for simplicity and in order to mainly concentrated on the effect of environmental parameters, we considered the minimum path loss rule in the backhaul links. A smart network operator may, however, use load balancing techniques to avoid congestion in the network, e.g., \cite{powall,loadj}. Finally, while we considered a fixed bandwidth split between the access and backhaul  which limits the resource allocation/coordination complexity, an adaptive split
between access and backhaul of different nodes would improve the network performance.}

\textcolor{black}{\subsubsection{On Some Practical Benefits of IAB}}
Using IAB with such a relatively small increment of the nodes reduces the network cost considerably\footnote{It is reasonable to consider almost the same cost for an IAB node and a typical SBS.}. This is because an SBS is much cheaper than fiber\footnote{Indeed, the exact cost of the fiber varies vastly in different regions, due to many factors including labour cost, etc. However, for different areas, fiber laying accounts to a significant fraction of the total network cost.}. 
For example, and only to give an intuitive view, as reported in \cite{ref8}, Table 7], in an urban area the fiber cost is estimated to be in the range of 20000 GBP/km, while an SBS in 5G is estimated to cost around 2500 GBP per unit \cite{ref11}\footnote{The price estimates are based on \cite{ref8}, and \cite{ref11}, and should not be considered as the cost estimations in Ericsson.}. More importantly, internal evaluations at Ericsson indicates that, for dense urban/\textcolor{black}{suburban} areas, even in the presence of dark fiber, the deployment of IAB networks is an opportunity to reduce the total cost of ownership (TCO) as well as the time-to-market. \textcolor{black}{Especially, the same hardware can be used both for access and backhaul such that no extra and seperate system is needed for backhaul.}

Thus, \textcolor{black}{although IAB may not} support the same peak rate as fiber, \textcolor{black}{IAB} will be \textcolor{black}{sufficient and} a cost-effective solution for \textcolor{black}{SBSs in} dense networks, \textcolor{black}{and} with no digging\footnote{According to different reports, e.g., \cite{ref9,ref10}, for fiber connection in metropolitan areas, a large portion (about 85\%) of the total \textcolor{black}{cost} figure is tied to trenching and installation.}, traffic jam and/or infrastructure displacement.

Along with the cost reduction, IAB increases the network flexibility remarkably. With optical fiber, the access points, of different types, can be installed only in the places with fiber connection. Such a constraint is, however, relaxed in IAB networks, and the nodes can be installed in different places as long as they have fairly good connection to their parent nodes. \textcolor{black}{These} are the reasons that different \textcolor{black}{operators} have shown interest to implement IAB in 5G networks \cite{ref12}, and \textcolor{black}{it is expected that IAB would be ultimately used} in up to 10-20\% of 5G sites\textcolor{black}{, e.g.,} \cite{ref13}.
\begin{figure}
\centerline{\includegraphics[width=3.5in]{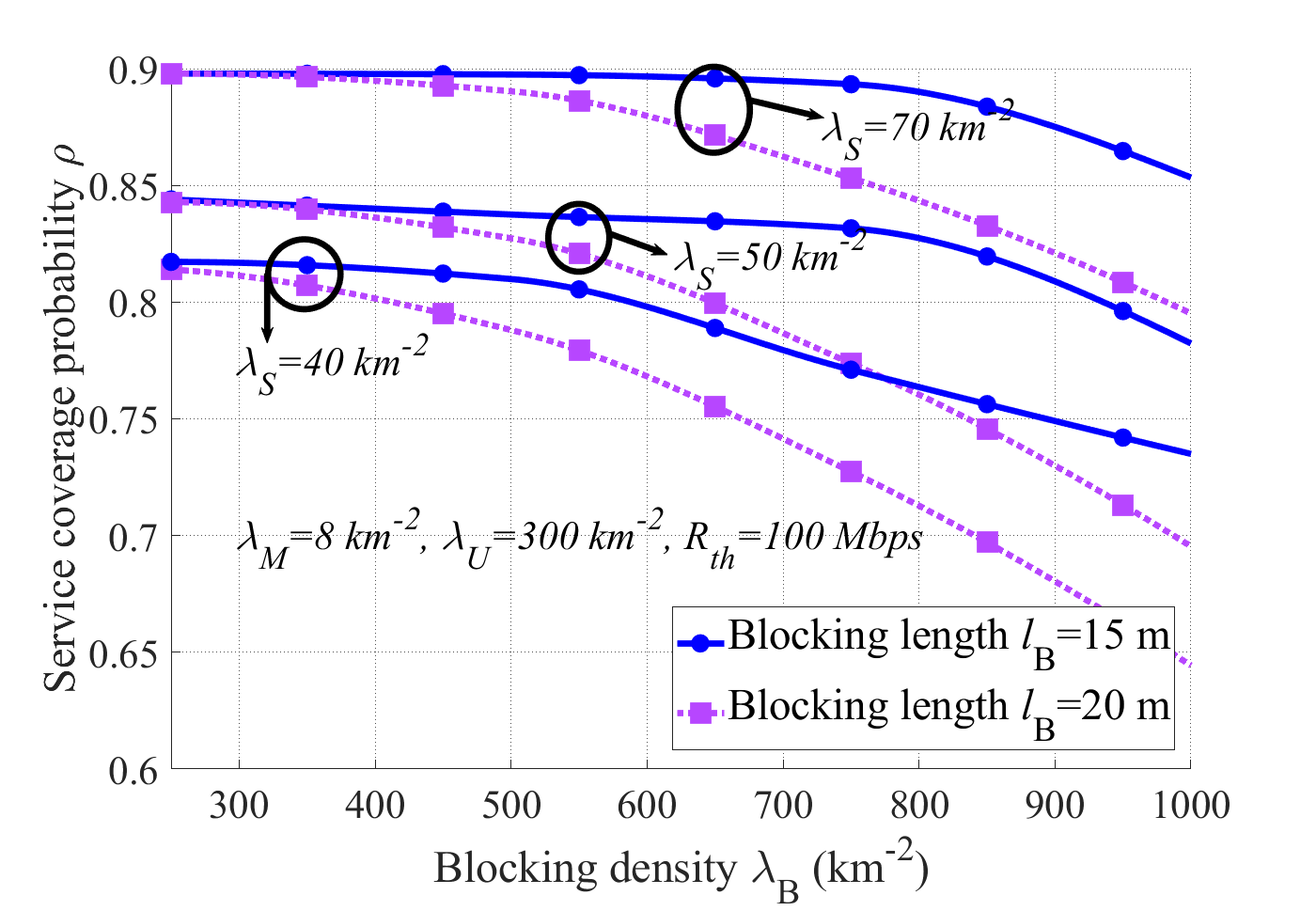}}
\caption{Service coverage probability of the IAB network as a function of the blocking density $\lambda_\text{B}$, with ${P_{\text{MBS}}, P_{\text{SBS}}, P_{\text{UE}}} = (40, 24, 0)$ dBm, and no rain/tree foliage.}
\label{blocking}
\end{figure}
     
It is interesting to note that, regardless of the cost, IAB is an attractive solution for a number of use-cases: 

\begin{itemize}
    \item Street trenching and digging not only are expensive but also may destroy historical areas or displace trees. For such reasons, some cities may consider a moratorium on fiber trenching \cite{ref9}, and instead rely on wireless backhaul methods such as IAB and \textcolor{black}{microwave backhaul}.  
    \item Fiber installation may take a long time, as it requires different permissions, labor work, etc. In such cases, IAB can establish new radio sites quickly. Thus, starting with IAB and, if/when needed, replacing it by fiber is expected to \textcolor{black}{become a} quite common setup.
    \item Low income zones of dense cities suffer from poor \textcolor{black}{Internet} connection. This is mainly because current fiber-based solutions are not economically viable, and the companies are not interested in fiber installation in such areas. Here, IAB is a low TCO solution to reduce the cost of \textcolor{black}{Internet} infrastructure.
    \item Public safety, and in general mission critical (MC), systems should be able to provide temporally on-demand coverage in all scenarios where the MC UEs are within terrestrial cellular network coverage or out of terrestrial cellular network coverage. In such cases, an IAB node, e.g., on a drone or a \textcolor{black}{fire truck}, can extend the coverage with high reliability and low latency\footnote{It should be noted that, within Rel-16 and 17, mobile IAB is not supported. Thus, with an IAB on, e.g., a drone, the node position should remain fixed during the data transmission.}.   
\end{itemize}
Finally, as expected and also emphasized in Fig. \ref{hybrid}, as the number of UEs increases, MBSs alone can not support the UEs' QoS requirements, and indeed we need to densify the network with, \textcolor{black}{e.g., using (IAB) nodes of different types.}

\subsubsection{Effect of Rain, Blocking and Tree Foliage}
\begin{figure}
\centerline{\includegraphics[width=3.5in]{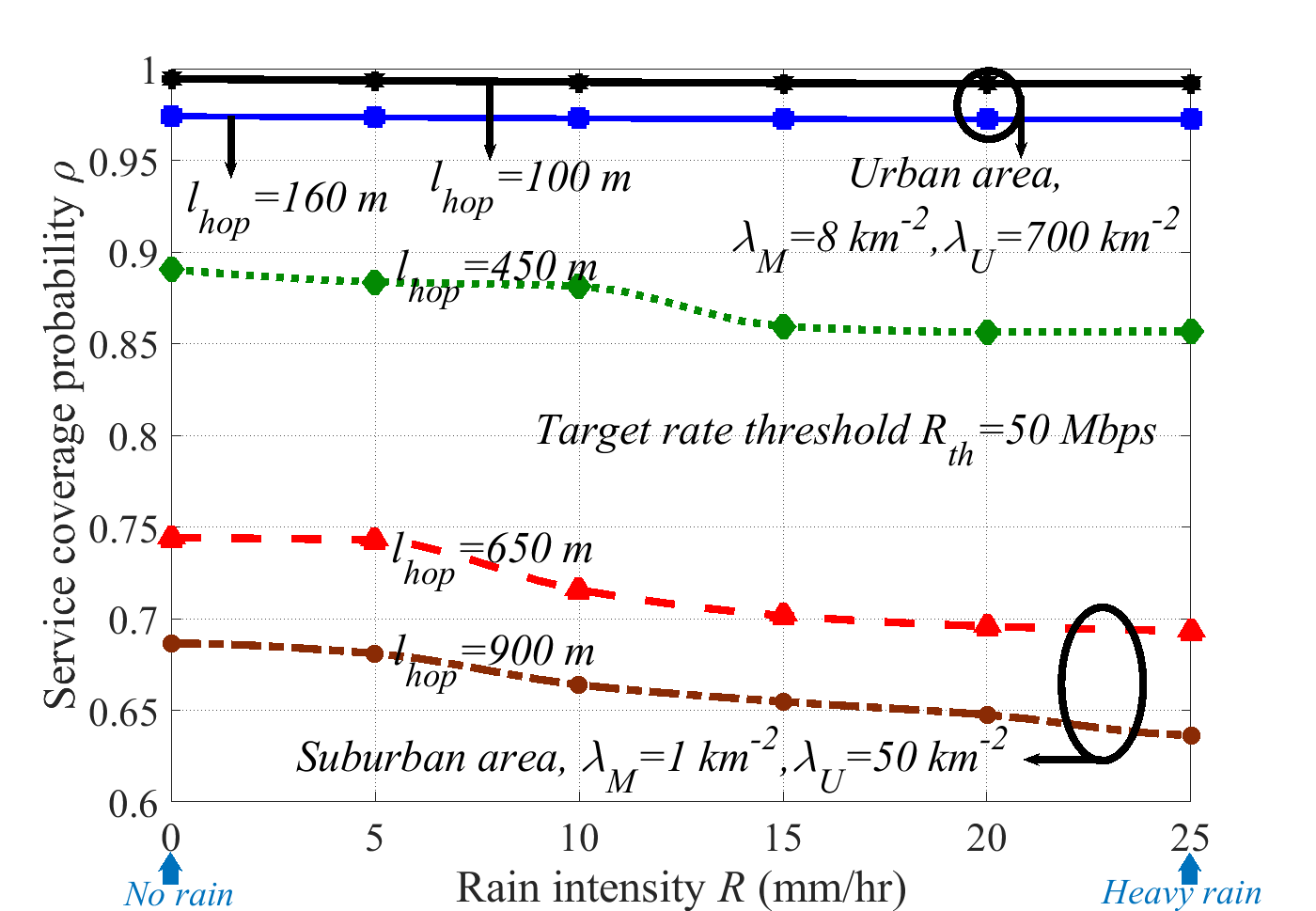}}
\caption{Service coverage probability of the IAB network as a function of the rain intensity in urban and \textcolor{black}{suburban} areas and for different average hop distances. The parameters are set to ${P_{\text{MBS}}, P_{\text{SBS}}, P_{\text{UE}}}$ = (45, 33, 0) dBm, $\lambda_{\text{B}}$ = 500 $\text{km}^{-2}$  for urban area and no blocking for the \textcolor{black}{suburban} area. Average hop distances $l_\text{hop}=100, 160, 450, 650, 900$ m correspond to SBS densities $\lambda_\text{S}$= 100, 50, 8, 5, 3 $\text{km}^{-2}$, respectively.}
\label{rain}
\end{figure}

\begin{figure}
\centerline{\includegraphics[width=3.5in]{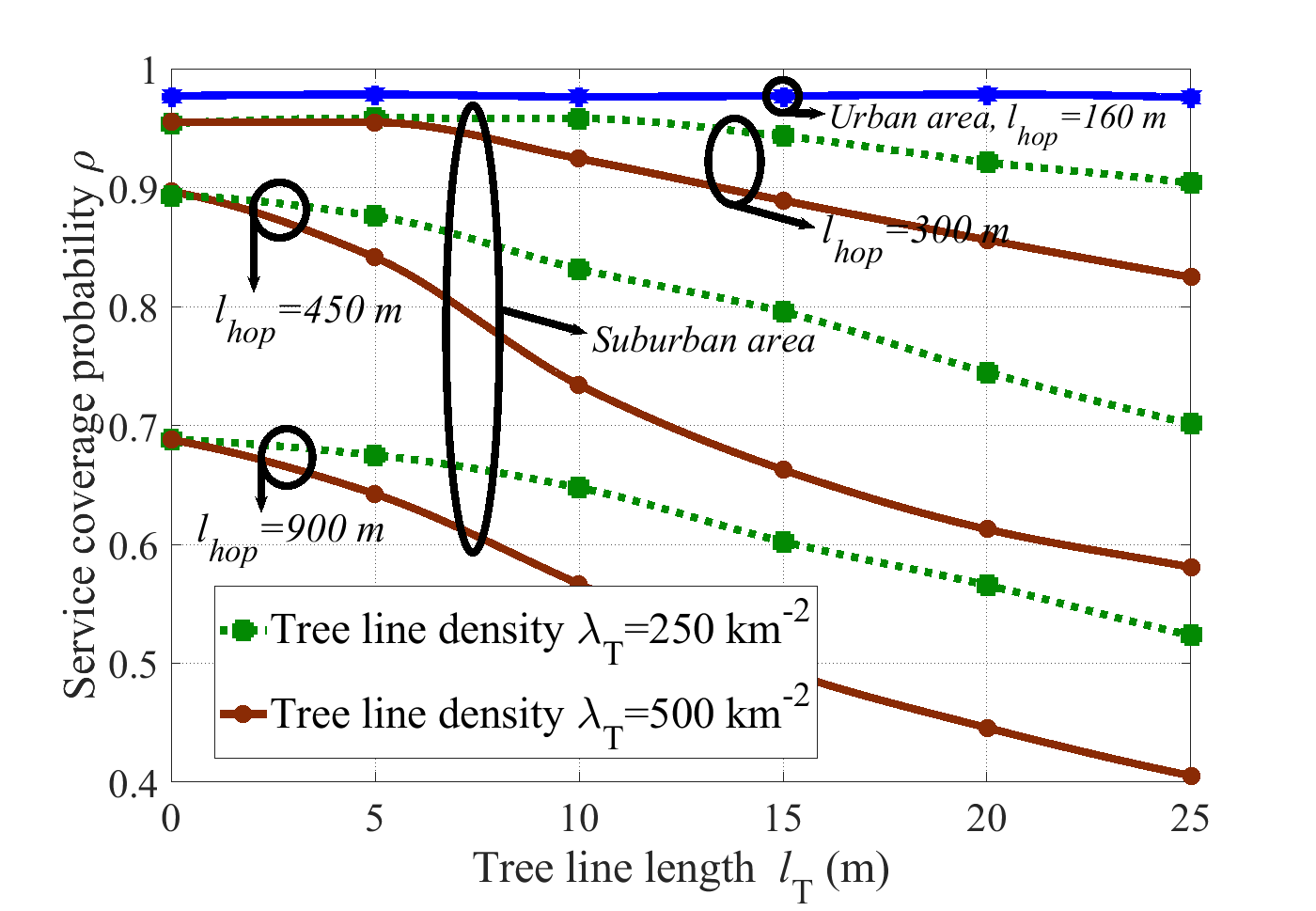}}
\caption{
Service coverage probability of the IAB network, in both \textcolor{black}{suburban} and urban areas, as a function of the tree line length $l_\text{T}$ with  $R_{\text{th}}$ = 50 Mbps, ${P_{\text{MBS}}, P_{\text{SBS}}, P_{\text{UE}}}$ = (45, 33, 0) dBm, $d$ = 5 m, in \eqref{eqvege}, and no rain. In the \textcolor{black}{suburban} area, we set $\lambda_\text{M}$ = 1 $\text{km}^{-2}$, $\lambda_\text{U}$ = 50 $\text{km}^{-2}$ with no blockage, while for the urban area we set $\lambda_\text{M}$ = 8 $\text{km}^{-2}$, $\lambda_\text{U}$ = 700 $\text{km}^{-2}$ with blockage having density $\lambda_\text{B}$ = 500$\text{km}^{-2}$ and length $l_\text{B}$ = 5 m. Average hop distances $l_\text{hop}$=160, 300, 450, 900 m correspond to SBS densities $\lambda_\text{S}$ = 50, 20, 8, 3 $\text{km}^{-2}$, respectively.
\label{vegetation}}
\end{figure}

As opposed to fiber-connected setups, an IAB network may be affected by \textcolor{black}{blockage, rain and tree foliage,} the effects of which are analyzed in Figs. \ref{blocking}-\ref{vegetation}, respectively. Particularly, considering the 2D FHPPP blockage model, Fig. \ref{blocking} investigates the coverage probability for different blockage densities $\lambda_\text{B}$ and walls lengths $l_\text{B}$ (also, see Fig. \ref{googleblock} for the effect of blockage in a 3D model).

Although IAB is of particular interest in \textcolor{black}{dense} urban areas, it has the potential to be used in \textcolor{black}{suburban} areas as well. For this reason, in Figs. \ref{rain} and \ref{vegetation} we demonstrate the coverage probability as a function of, respectively, the rain intensity, $R$ in (6), and the tree line length $l_\text{T}$ in both urban and \textcolor{black}{suburban} areas. Here, the results are presented for the average hop distances $l_\text{hop}$=100, 160, 450, 650 ,900 m which correspond to SBSs densities $\lambda_\text{S}$=100, 50, 8, 5 and 3 $\text{km}^{-2}$, respectively. For a \textcolor{black}{suburban} area, i.e., the cases with large average hop distance, we consider a single MBS, no blockage and UEs’ density $\lambda_\text{U}=$50 $\text{km}^{-2}$. On the other hand, for the cases with urban areas, i.e., low average hop distance, the blockage and the UEs, densities are set to $\lambda_\text{B}=500 \text{ km}^{-2}$ and $\lambda_\text{U}=700 \text{ km}^{-2}$, respectively. According to Figs. \ref{blocking}-\ref{vegetation}, the following points can be concluded:
\begin{figure}
\centerline{\includegraphics[width=3.5in]{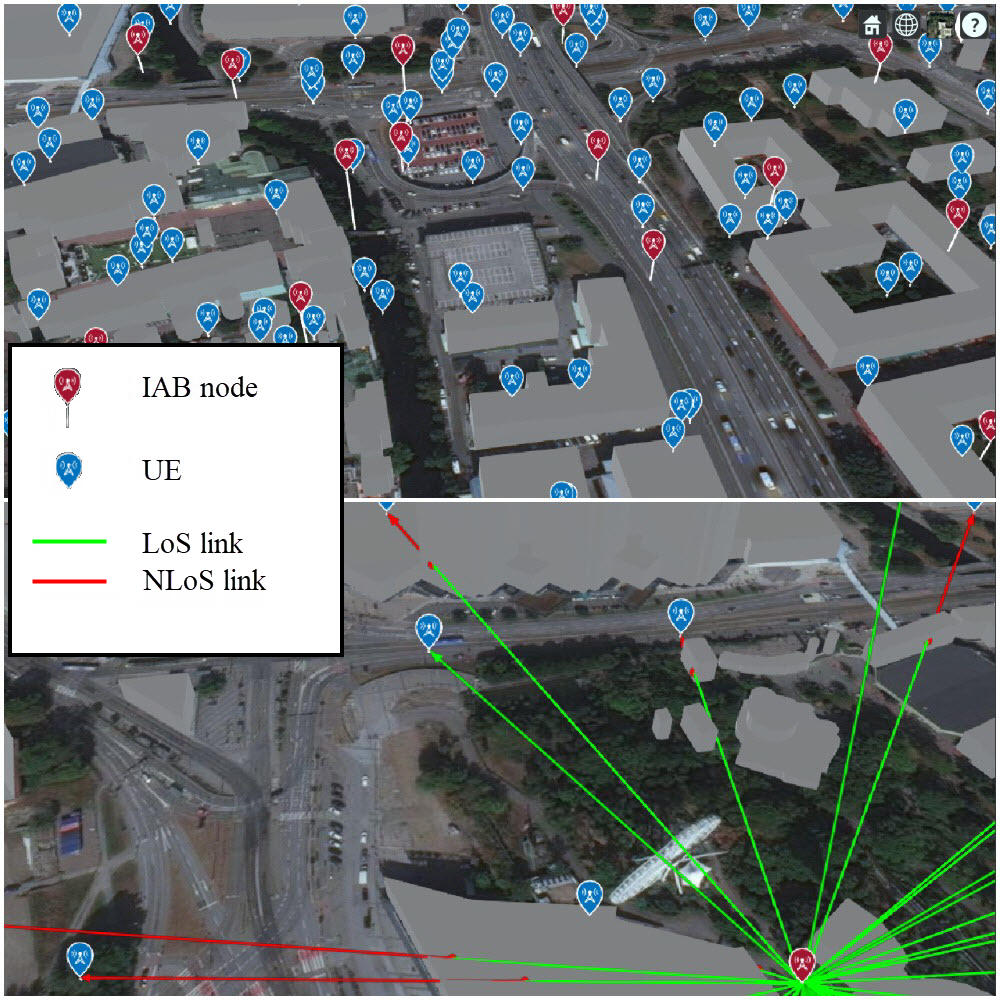}}
\caption{An example of the distribution of the IAB network in 3D space with OpenStreetMap.
\label{figgoogle}}
\end{figure}

\begin{itemize}
   
\item Unless for low network densities, the coverage probability is not \textcolor{black}{much} affected by the blockage density/length (Fig. \ref{blocking}. \textcolor{black}{also}, see Fig. \ref{googleblock} for the effect of blockage in a 3D example use-case). This is intuitive because, as the network density increases, with high probability each UE can be connected to an SBS with strong LOS signal component.

\item Considering 28 GHz, rain will not be a problem for IAB, unless for the cases with very heavy \textcolor{black}{rainfall} in \textcolor{black}{suburban} areas (Fig. \ref{rain}). Particularly, the system performance is robust to different rain intensities in suburban/urban areas. Moreover, in \textcolor{black}{suburban} areas, even with high intensities the rain reduces the coverage probability slightly\footnote{It should be noted that, while Fig. \ref{rain} presents the results for 28 GHz which is the frequency of interest for IAB, the effect of the rain will be more visible at higher carrier frequencies.}.

\item As opposed to the rain and the blockage, depending on the network density, in the cases with low/moderate IABs' densities the coverage probability may be considerably affected by the tree foliage. For instance, consider the parameter settings of Fig. \ref{vegetation}, \textcolor{black}{in suburban area,} with 1 MBS, $\lambda_\text{U}=50 \text{ km}^{-2}$ and an average hop distance of $l_\text{hop}=900$ m, corresponding to $\lambda_\text{S}=3 \text{ km}^{-2}$. Then, the presence of trees with line length $l_\text{T}=15$ m and density $\lambda_\text{T}=250 \text{ km}^{-2}$ reduces the coverage probability from $70\%$ for the cases with no trees to $60\%$, i.e., for $10\%$ more of the UEs the rate requirement $50$ Mbps can not be provided. Thus, in the presence of tree foliage, more IAB nodes are required to satisfy the same QoS requirement. On the other hand, with high network density, the coverage probability is not affected by the tree foliage (Fig. 9). In general, predicting the link performance for IAB is difficult when accepting foliage. This is because, for instance, the backhaul link quality may change due to wet trees, snow on the trees, wind and varying percentage of leaves in different seasons. However, based on the presented results, we believe \textcolor{black}{that, with appropriate nodes heights,} mmWave IAB will work well for areas with low/moderate foliage level.

\begin{figure}
\centerline{\includegraphics[width=3.5in]{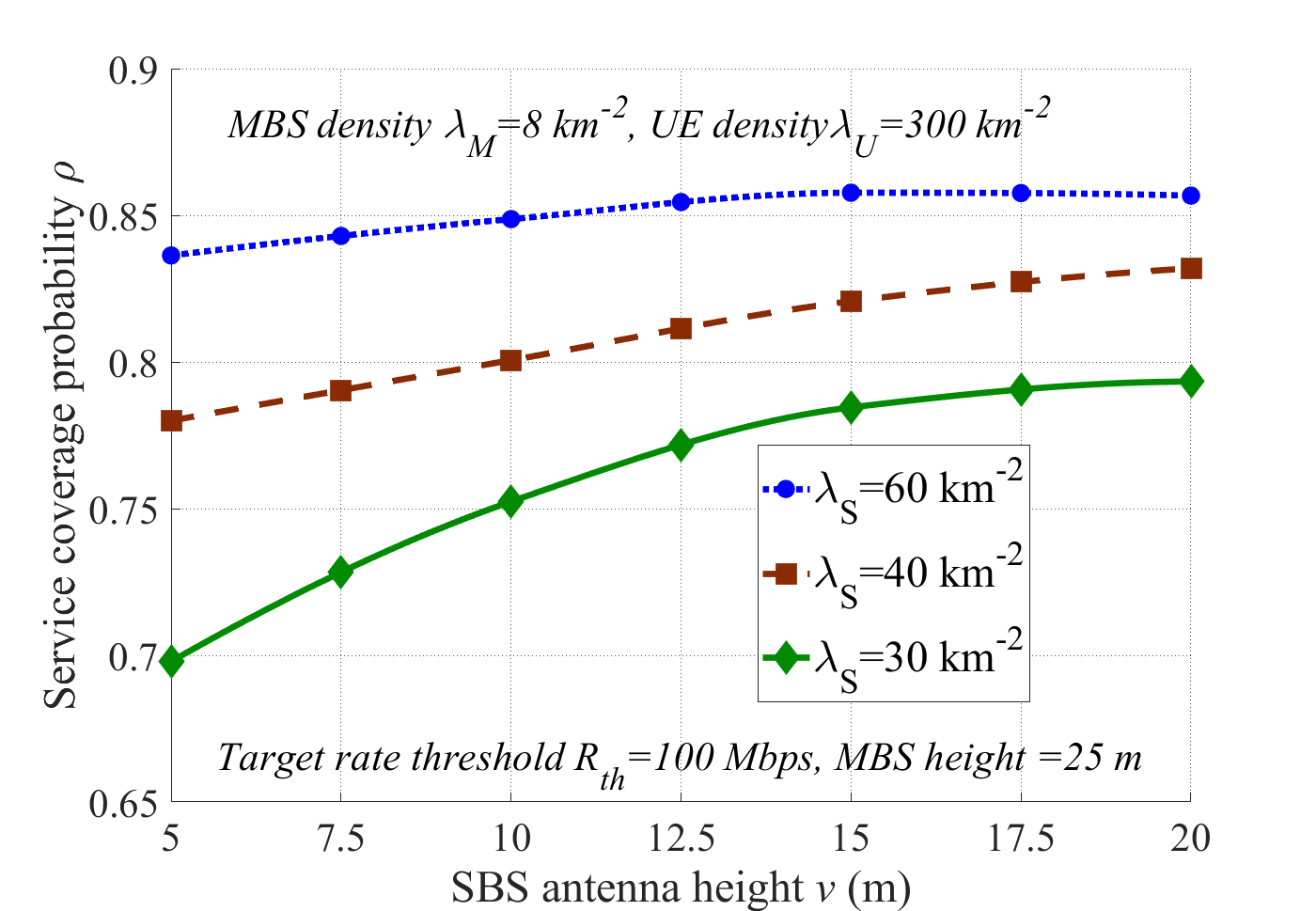}}
\caption{Service coverage probability as a function of the SBSs antenna height $v$ for the cases with no rain and ${P_{\text{MBS}}, P_{\text{SBS}}, P_{\text{UE}}} = (40, 24, 0)$ dBm.\label{googleblock}}
\end{figure}

\end{itemize}
Finally, it should be mentioned that in Figs. \ref{rain}-\ref{vegetation} we considered the same parameter settings for the IAB nodes, independently of their area of implementation. However, in practice, different types of short-range and wide-area IAB nodes, with considerably higher capabilities for the wide-area IAB nodes, may be developed and used in urban and \textcolor{black}{suburban} areas, respectively \cite{ref52}. This will help to reduce the effect of rain/foliage in \textcolor{black}{suburban} areas even more.


\subsubsection{Performance Evaluation in an Example 3D Use-case}\label{3d}
\begin{figure}
\centerline{\includegraphics[width=3.5in]{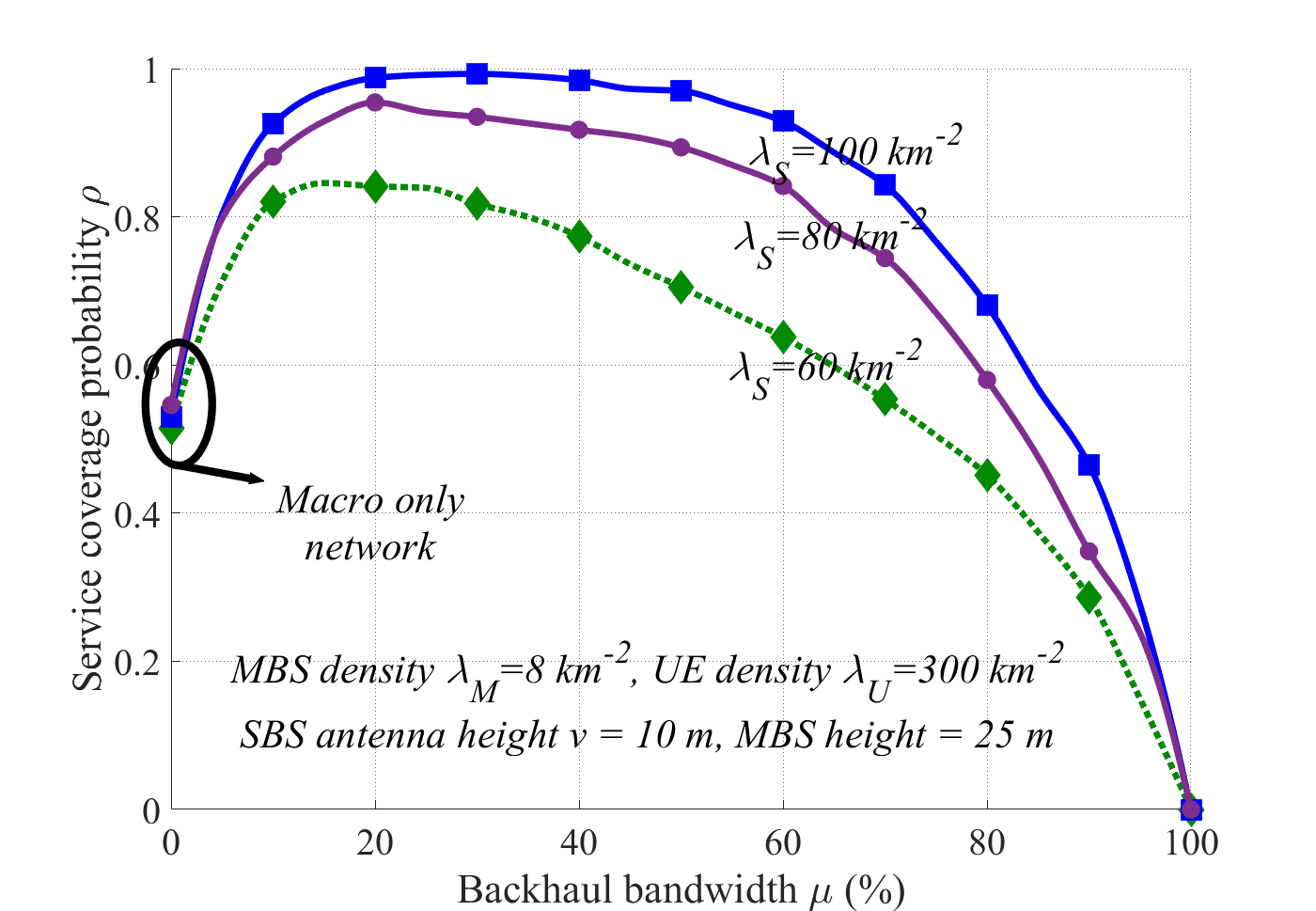}}
\caption{Service coverage probability as a function of the backhaul bandwidth allocation percentage $\mu$ in \eqref{eq:11} for a dense network, no rain, $R_{\text{th}}$ = 100 Mbps, and ${P_{\text{MBS}}, P_{\text{SBS}}, P_{\text{UE}}} = (40, 24, 0)$ dBm.\label{partition}}
\end{figure}

In Figs. \ref{iabeq}-\ref{vegetation}, we \textcolor{black}{investigate} the system performance in the 2D FHPPP-based model. To evaluate the effect of the nodes and blockages heights, in this subsection we study the coverage probability in an example 3D setup. Particularly, as shown in Fig. \ref{figgoogle}, the UEs and the IAB nodes (both MBSs and SBSs) are still randomly distributed based on their corresponding FHPPPs, while they are positioned at different heights. Moreover, the blockages (as well as the distance between the nodes) are determined based on the map information, i.e., the real world blocking terrain is considered using OpenStreetMap 3D environment. The results have been tested on a disk of radius $D= 0.5 \text{ km}$ over the Chalmers University of Technology, Gothenburg, Sweden. Particularly, considering the MBSs and the UEs heights to be 25 and 1 m, respectively, Figs. \ref{googleblock} and \ref{partition} show the coverage probability as a function of the SBSs’ heights and the backhaul bandwidth allocation percentage, $\mu$ in \eqref{eq:11}, respectively.  

As demonstrated in Fig. \ref{googleblock}, with a low SBS density, increasing the height of the SBSs helps to reduce the required number of IAB nodes considerably. For instance, with the parameter settings of Fig. \ref{googleblock}, the same coverage probability as in the cases with density $\lambda_\text{S}=40 \text{ km}^{-2}$ and height $v=5 $ m is achieved by a setup having $\lambda_\text{S}=30 \text{ km}^{-2}$ and $v=15 $ m. However, as the network density increases, the effect of the SBSs height becomes negligible. This is intuitively because, with moderate/high densities, with high probability one can always find IAB donor-IAB, IAB-UE, and IAB donor-UE links with strong LOS signal components, even if the IAB nodes are located on the street level. 

Finally, as shown in Fig. \ref{partition}, with an optimal bandwidth allocation between the access and backhaul, IAB network increases the coverage probability, compared to the cases with only MBSs, significantly (Also, see Fig. \ref{hybrid}). With $\mu=0,$ the system performance decreases to those achieved by only MBSs, as no bandwidth is allocated for backhauling. With $\mu=100\%,$ on the other hand, no resources are considered for access, and the coverage probability tends to zero. Thus, for different parameter settings, there is an optimal value for the portion of backhaul/access bandwidth allocation maximizing the coverage probability \textcolor{black}{(Fig. \ref{partition})}. Deriving this optimal value, which increases with the SBSs’ density and decreases with the UEs’ density, is an open research topic for which the results of \cite{refi17}  is supportive.

\section{Conclusion}\label{conclusion}
We studied IAB networks from both standardization and performance points of view. As we showed, depending on the QoS requirements, IAB can be considered as a cost-effective \textcolor{black}{alternative to optical fiber that complements conventional microwave backhaul}, in different use-cases and areas. Particularly, the same coverage probability as in fiber-connected networks is achieved by relatively small increment in the number of IAB nodes, leading to considerable network cost reduction/flexibility increment. Moreover, unless for the cases with moderate/high tree foliage in \textcolor{black}{suburban} areas, the system performance is not much affected by, e.g., the blockage, the rain, and the tree foliage, which introduces the IAB as a robust setup for dense networks.

While the industry has well proceeded in standardization of different aspects of the network, there are still many open research problems to be addressed by the academia. Among such research topics are topology optimization using, e.g., machine learning, studying the effect of hardware impairments on the system performance, developing efficient methods for simultaneous transmission/reception, improving the system performance using network coding, designing efficient (hybrid) beamforming methods for IAB networks, \textcolor{black}{combination of IAB nodes and repeaters/intelligent surfaces,} as well as mobile IAB. \textcolor{black}{Also, load balancing and adaptive routing in a mesh-based network are interesting research topics for which the fundamental results of relay networks, e.g., \cite{referencex_new,referencey_new,referencez_new}, will be supportive.} Although some of these topics are not supported in Rel-16 and 17, a deep analysis of such problems may pave the way for further enhancements of IAB in industry.

\bibliographystyle{IEEEtran}
\bibliography{bibliography}

\begin{IEEEbiography}[{\includegraphics[width=1in,clip,keepaspectratio]{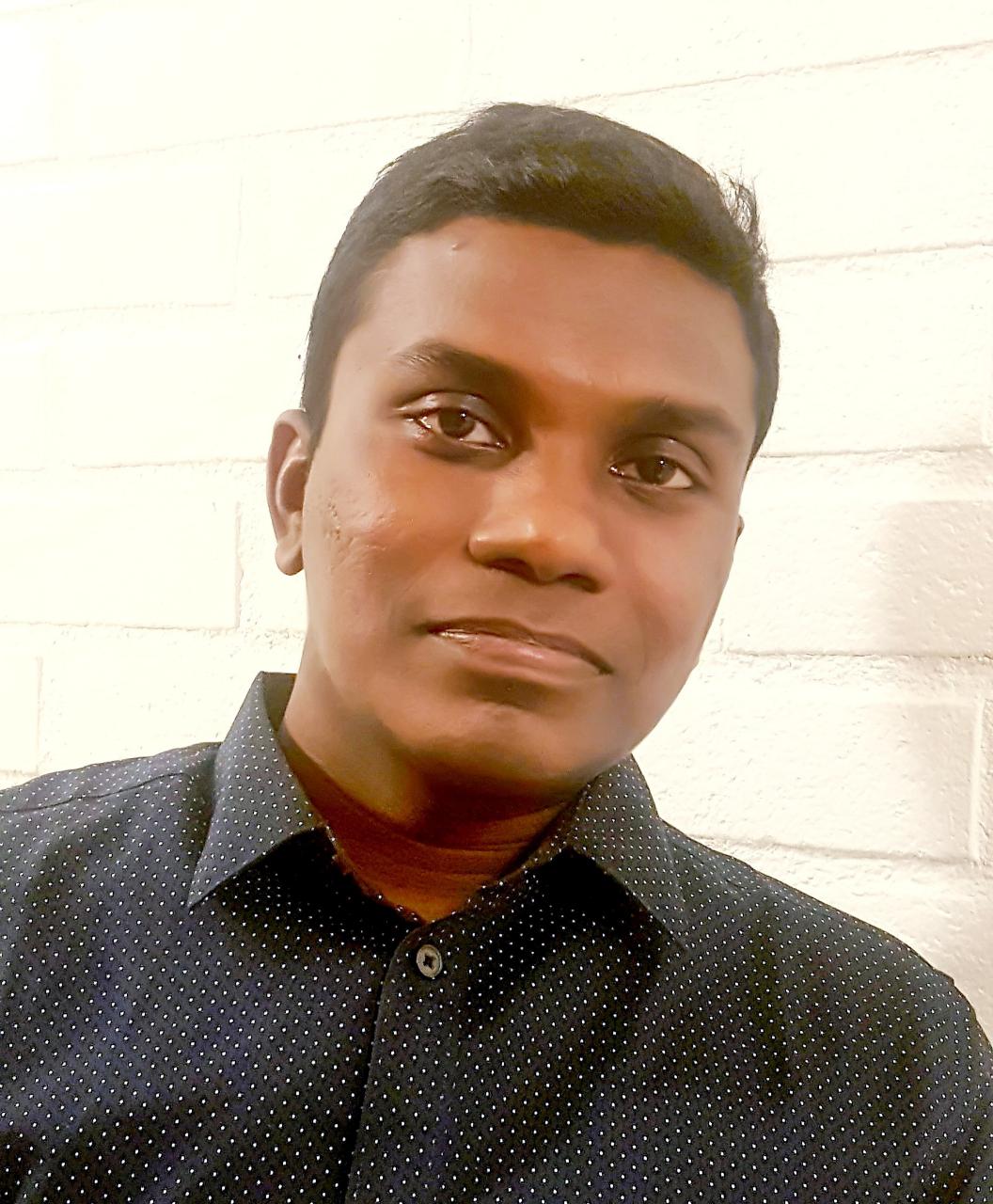}}]{Charitha Madapatha} \textcolor{black}{is currently pursuing the Ph.D. in Communication Systems and Information Theory from Chalmers University of Technology, Gothenburg, Sweden. He received his MSc. degree in Communication Engineering from Chalmers in 2020 and the BSc. degree in Telecommuncations Engineering from Asian Institute of Technology, Pathumthani, Thailand, in 2016. }

Charitha is the recipient of Swedish Institute Scholarship for Global Professionals, Sweden, 2018 and the AIT Fellowship grant, Thailand, 2014-2016. He was also involved in LTE network planning and optimization in Thailand and Cambodia. His current research interests include integrated access and backhaul, multi antenna systems, mmWave communication, analysis of physical layer algorithms, resource allocation. He has co-authored several international scientific publications in the field of wireless networks.

\end{IEEEbiography}

\begin{IEEEbiography}[{\includegraphics[width=1in,clip,keepaspectratio]{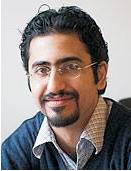}}]{Behrooz Makki} [M'19, SM'19] received his PhD degree in Communication Engineering from Chalmers University of Technology, Gothenburg, Sweden. In 2013-2017, he was a Postdoc researcher at Chalmers University. Currently, he works as Senior Researcher in Ericsson Research, Gothenburg, Sweden.

Behrooz is the recipient of the VR Research Link grant, Sweden, 2014, the Ericsson's Research grant, Sweden, 2013, 2014 and 2015, the ICT SEED grant, Sweden, 2017, as well as the Wallenbergs research grant, Sweden, 2018. Also, Behrooz is the recipient of the IEEE best reviewer award, IEEE Transactions on Wireless Communications, 2018. \textcolor{black}{Currently, he works as an Editor in IEEE Wireless Communications Letters, IEEE Communications Letters, the journal of Communications and Information Networks, as well as the Associate Editor in Frontiers in Communications and Networks.} He was a member of European Commission  projects ``mm-Wave based Mobile Radio Access Network for 5G Integrated Communications'' and ``ARTIST4G'' as well as various national and international research collaborations. His current research interests include integrated access and backhaul, hybrid automatic repeat request, Green communications, millimeter wave communications, free-space optical communication, NOMA, finite block-length analysis and backhauling. He has co-authored 57 journal papers, 45 conference papers and 40 patent applications.

\end{IEEEbiography}
\begin{IEEEbiography}[{\includegraphics[width=1in,clip,keepaspectratio]{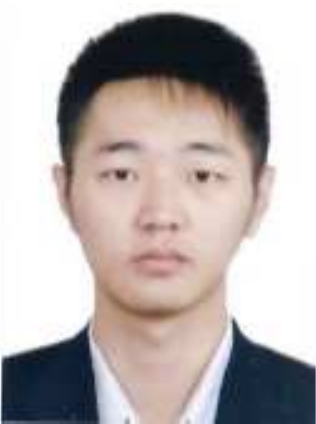}}]{Chao Fang} received the B.E. degree in Communication Engineering from Hubei University, Wuhan, China, in 2013, and the M.S. degree in Electrical Engineering from Chalmers University of Technology, Gothenburg, Sweden, in 2015, where he is currently pursuing the Ph.D. degree with the Communication Systems Group, Department of Electrical Engineering. His research interests include beamforming, millimeter-wave communication, integrated access backhauled networks, heterogeneous cellular networks and stochastic geometry.

\end{IEEEbiography}
\begin{IEEEbiography}[{\includegraphics[width=1in,clip,keepaspectratio]{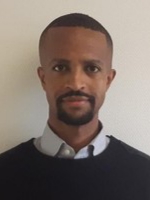}}]{Dr. Oumer Teyeb} is currently a master researcher at Ericsson Research, Sweden. He earned a Ph.D. in mobile communications from Aalborg University, Denmark, in 2007 and has been working at Ericsson Research in Stockholm, Sweden, since 2011. His main areas of research areas are protocol and the architectural aspects of cellular networks, and the interworking of cellular networks with local area wireless networks such as WLAN. Since 2017, he has been part of the Ericsson 3GPP standardization delegation in the RAN2 WG. He is the inventor/co-inventor of 250+ patent families, as well as the author/co-author several international scientific publications and standardization contributions in the field of wireless networks.

\end{IEEEbiography}
\begin{IEEEbiography}[{\includegraphics[width=1in,clip,keepaspectratio]{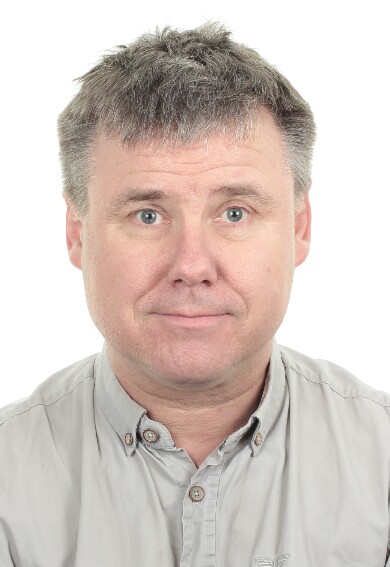}}]{Dr. Erik Dahlman} is currently Senior Expert in Radio Access Technologies within Ericsson Research. He has been deeply involved in the development of all 3GPP wireless access technologies, from the early 3G technologies (WCDMA(HSPA), via 4G LTE, and most recently the 5G NR technology. He current work is primarily focusing on the evolution of 5G as well as technologies applicable to future beyond 5G wireless access.

Erik Dahlman is the co-author of the books 3G Evolution – HSPA and LTE for Mobile Broadband, 4G – LTE and LTE-Advanced for mobile broadband, 4G – LTE-Advanced Pro and The Road to 5G and, most recently, 5G NR – The Next Generation Wireless Access Technology.

In 2009, Erik Dahlman received the Major Technical Award, an award handed out by the Swedish Government, for his contributions to the technical and commercial success of the 3G HSPA radio-access technology. In 2010, he was part of the Ericsson team receiving the LTE Award for “Best Contribution to LTE Standards”, handed out at the LTE World Summit. In 2014 he was nominated for the European Inventor Award, the most prestigious inventor award in Europe, for contributions to the development of 4G LTE.
\end{IEEEbiography}
\begin{IEEEbiography}[{\includegraphics[width=1in,clip,keepaspectratio]{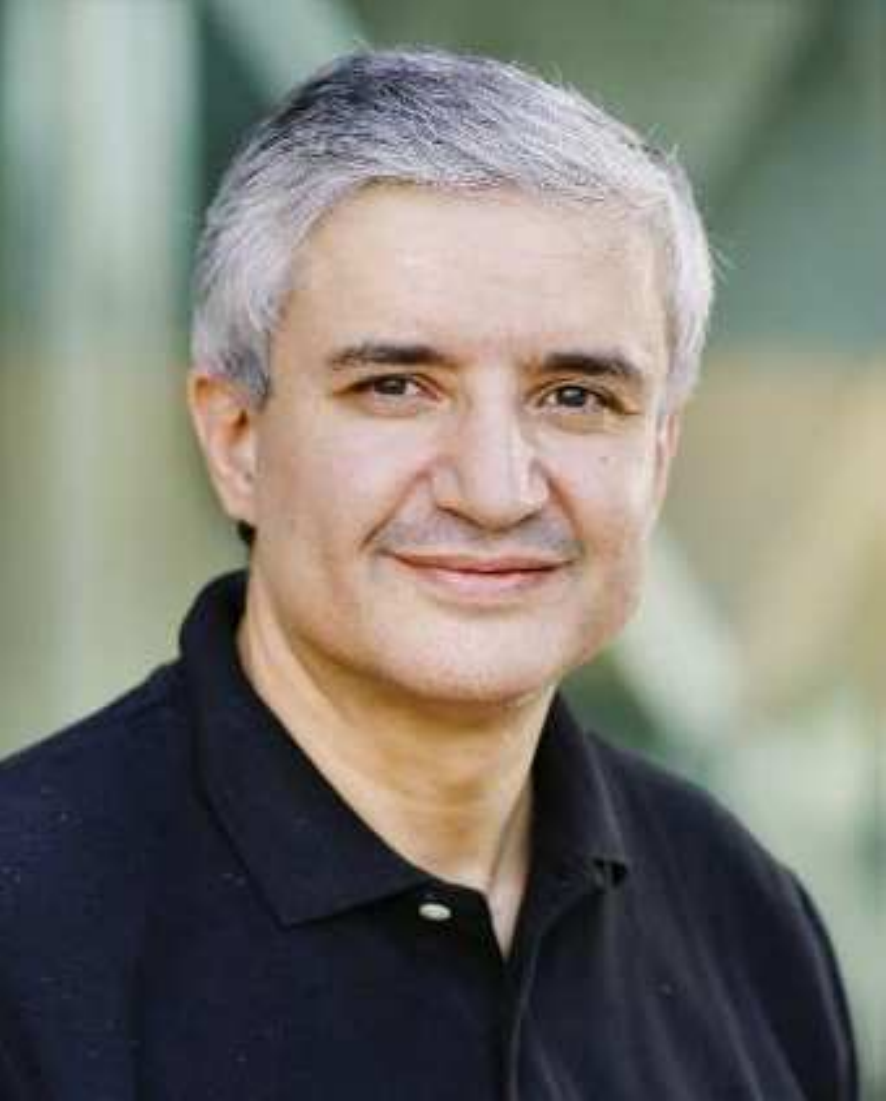}}]{Mohamed-Slim Alouini} (S'94, M'98, SM'03, F'09) was born in Tunis, Tunisia. He received the Ph.D. degree in Electrical Engineering from the California Institute of Technology (Caltech), Pasadena, CA, USA, in 1998. He served as a faculty member in the University of Minnesota, Minneapolis, MN, USA, then in the Texas A$\&$M University at Qatar, Education City, Doha, Qatar before joining King Abdullah University of Science and Technology (KAUST), Thuwal, Makkah Province, Saudi Arabia as a Professor of Electrical Engineering in 2009. His current research interests include the modeling, design, and performance analysis of wireless communication systems.
\end{IEEEbiography}
\begin{IEEEbiography}[{\includegraphics[width=1in,clip,keepaspectratio]{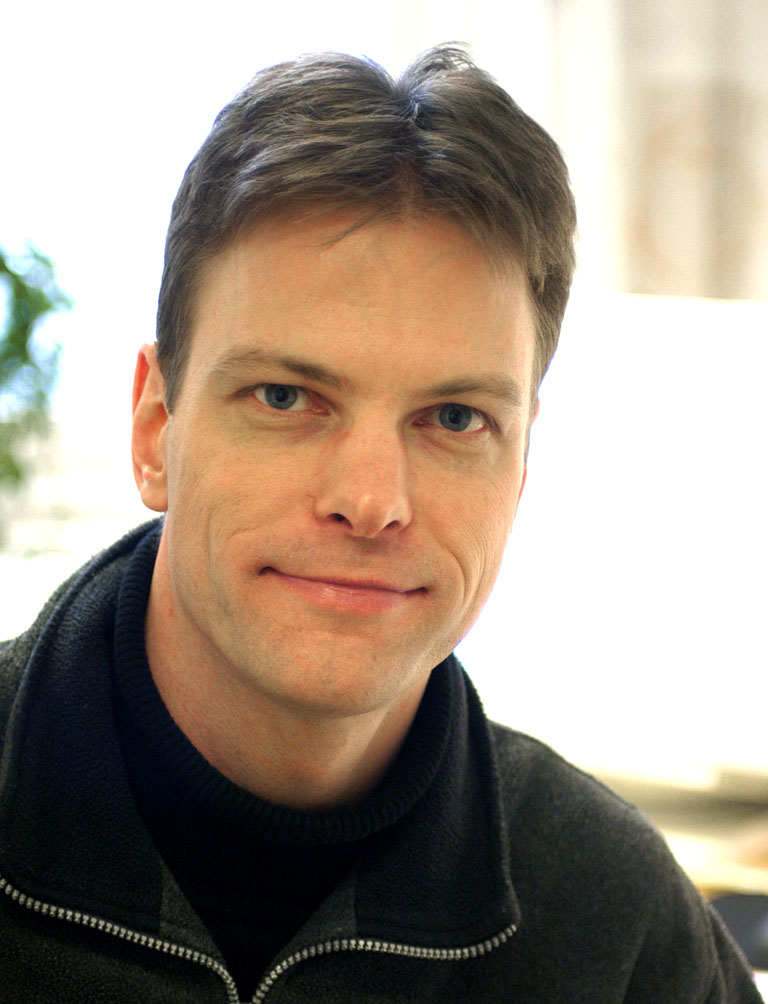}}]{Tommy Svensson} [S’98, M’03, SM’10] is Full Professor in Communication Systems at Chalmers University of Technology in Gothenburg, Sweden, where he is leading the Wireless Systems research on air interface and wireless backhaul networking technologies for future wireless systems. He received a Ph.D. in Information theory from Chalmers in 2003, and he has worked at Ericsson AB with core networks, radio access networks, and microwave transmission products. He was involved in the European WINNER and ARTIST4G projects that made important contributions to the 3GPP LTE standards, the EU FP7 METIS and the EU H2020 5GPPP mmMAGIC and 5GCar projects towards 5G and beyond, as well as in the ChaseOn antenna systems excellence center at Chalmers targeting mm-wave solutions for 5G access, backhaul/ fronthaul and V2X scenarios. His research interests include design and analysis of physical layer algorithms, multiple access, resource allocation, cooperative systems, moving networks, and satellite networks. \textcolor{black}{He has co-authored 4 books, 86 journal papers, 126 conference papers and 53 public EU projects deliverables. He is Chairman of the IEEE Sweden joint Vehicular Technology/ Communications/ Information Theory Societies chapter, founding editorial board member and editor of IEEE JSAC Series on Machine Learning in Communications and Networks, has been editor of IEEE Transactions on Wireless Communications, IEEE Wireless Communications Letters, Guest editor of several top journals, organized several tutorials and workshops at top IEEE conferences, and served as coordinator of the Communication Engineering Master's Program at Chalmers.
}

\end{IEEEbiography}

\end{document}